\documentclass[12pt,a4paper]{article}
\usepackage{graphicx}
\usepackage{amssymb}
\usepackage{amsmath}
\usepackage{bm}
\usepackage{color}
\usepackage{tikz}
\usepackage{theorem}
\usepackage{cite}
\usepackage{amsfonts}
\usepackage{multirow}
\usepackage{changepage}
\usepackage{booktabs}

\usepackage{graphicx}
\usepackage{caption}
\usepackage{subcaption}
\usepackage{nicematrix}

\usepackage{geometry}
\usepackage[compat=1.0.0]{tikz-feynman}

\usepackage[sort&compress,numbers, merge]{natbib}

\newcommand{\eprint}[2][]{\href{https://arxiv.org/abs/#2}{arXiv:~\nolinkurl{#2}}}

\usepackage{scrextend}

\definecolor{Orange}{cmyk}{0,0.61,0.87,0}
\definecolor{JungleGreen}{cmyk}{0.99,0,0.52,0} 
\definecolor{OliveGreen}{cmyk}{0.64,0,0.95,0.40}
\definecolor{Brown}{cmyk}{0,0.81,1,0.60}
\definecolor{RoyalBlue}{cmyk}{0.71,0.53,0,0.12}
\definecolor{darkspringgreen}{rgb}{0.09, 0.45, 0.27}

\newcommand{\be}{\begin{equation}}
\newcommand{\ee}{\end{equation}}
\newcommand{\bea}{\begin{eqnarray}}
\newcommand{\eea}{\end{eqnarray}}
\newcommand{\eq}[1]{Eq.~(\ref{#1})}

\usetikzlibrary{decorations.markings,angles,quotes,positioning,decorations.pathreplacing,decorations.markings,snakes,arrows,calc}
\usetikzlibrary{fadings}
\usetikzlibrary{shapes.misc}
\tikzset{crossr/.style={cross out, draw=red, minimum size=4*(#1-\pgflinewidth), inner sep=0pt, outer sep=0pt},
crossr/.default={2pt}}
\tikzset{crossb/.style={cross out, draw=black, minimum size=4*(#1-\pgflinewidth), inner sep=0pt, outer sep=0pt},
crossb/.default={2pt}}
\tikzset{crossp/.style={cross out, draw=violet, minimum size=4*(#1-\pgflinewidth), inner sep=0pt, outer sep=0pt},
crossp/.default={2pt}}

\tikzfading 
[
  name=fade out,
  inner color=transparent!0,
  outer color=transparent!100
]

\tikzset{
upp/.style={postaction={decorate},
  decoration={markings,mark=at position .5 with  \arrow{>}}},
  dow/.style={postaction={decorate},
  decoration={markings,mark=at position .5 with  \arrow{<}}},
 }

\tikzset{
    ncbar angle/.initial=90,
    ncbar/.style={
        to path=(\tikztostart)
        -- ($(\tikztostart)!#1!\pgfkeysvalueof{/tikz/ncbar angle}:(\tikztotarget)$)
        -- ($(\tikztotarget)!($(\tikztostart)!#1!\pgfkeysvalueof{/tikz/ncbar angle}:(\tikztotarget)$)!\pgfkeysvalueof{/tikz/ncbar angle}:(\tikztostart)$)
        -- (\tikztotarget)
    },
    ncbar/.default=0.5cm,
}

\tikzset{round left paren/.style={ncbar=0.5cm,out=100,in=-100}}
\tikzset{round right paren/.style={ncbar=0.5cm,out=80,in=-80}}

\usepackage[colorlinks=True, citecolor=blue, linkcolor=blue, urlcolor=black]{hyperref}

\allowdisplaybreaks[1]

\renewcommand{\arraystretch}{1.3}

\begin{document}

\begin{titlepage}
\begin{center}
\hfill

\vspace{2.0cm}
{\Large\bf  
Bootstrapping the Chiral-Gravitational Anomaly
}

\vspace{2cm}
{\bf 
Zi-Yu Dong$^1$, Teng Ma$^{2}$, Alex Pomarol$^{1,3}$ and Francesco Sciotti$^1$
}
\\
\vspace{0.7cm}
{\it\footnotesize
${}^1$IFAE and BIST, Universitat Aut\`onoma de Barcelona, 08193 Bellaterra, Barcelona\\
${}^2$ ICTP-AP, 
University of Chinese Academy of Sciences, 100190 Beijing, China\\
${}^3$Departament de F\'isica, Universitat Aut\`onoma de Barcelona, 08193 Bellaterra, Barcelona\\
}

\vspace{0.9cm}
\abstract
We  analyze  causality and unitarity constraints
in graviton  scattering amplitudes, aiming 
to  establish new bounds on theories with $U(1)$-gravitational anomalies, such as axion models
or strongly-coupled gauge theories. For this purpose, 
we show the necessity  of coupling these theories to  gravity.
We obtain a universal scale  $\Lambda_{\rm caus}$ at which
 states with $J\geq 4$ must appear in the theory. 
 We show that  this scale can lie below the quantum gravity scale. 
 For axion models, we get  $\Lambda_{\rm caus}\sim\sqrt{M_P f_a}$ where $f_a$ is the axion decay constant.
In strongly-coupled gauge theories in the  large-$N_c$ limit,
 the presence of glueballs  allows to 
  evade these bounds,  provided the  number of fermions $N_F\ll N_c$ and the 'tHooft coupling is not  large. 
Nevertheless, for models that have a holographic 5D dual (large 'tHooft coupling),
$\Lambda_{\rm caus}$  
emerges as a new cutoff scale, unless certain conditions on the parameters of the 5D models are satisfied.

\end{center}
\end{titlepage}
\setcounter{footnote}{0}

\section{Introduction}

Bootstrap techniques provide a powerful way to constrain quantum field theories \cite{Kruczenski:2022lot}.
This approach has been successfully applied to  bound
strongly-coupled gauge theories like  QCD \cite{Paulos:2016but, Paulos:2017fhb, Doroud:2018szp,Guerrieri:2018uew, Guerrieri:2020bto, Zahed:2021fkp,Alvarez:2021kpq, EliasMiro:2022xaa, He:2023lyy, Acanfora:2023axz, Guerrieri:2023qbg, He:2024nwd, Guerrieri:2024jkn},
based on the requirement of  analyticity, unitarity and crossing  in scattering amplitudes.

Recently, it has also been used to    efficiently corner
gauge theories  in the large number of "colors"  limit ($N_c\to\infty$)
\cite{ Albert:2022oes,Fernandez:2022kzi,Albert:2023jtd,Ma:2023vgc,Li:2023qzs,Albert:2023seb}
where amplitudes can be  simply described by  the tree-level exchange of a tower of resonances.
The  initial analysis from  pion-pion scattering \cite{Albert:2022oes}  
provided already a classification of all  possible  UV completions for the effective field theory (EFT)  of pions \cite{Albert:2022oes,Fernandez:2022kzi,Li:2023qzs,Albert:2023seb}.
These could be separated in 3  groups:
 models with scalar resonances (like Higgs models), models with  $J=1$ states (as 5D holographic models),  and models of higher-spin resonances 
that range from models with   $J=1,2,...$ degenerate states ($su$-model) to stringy models (such as the Veneziano or Lovelace-Shapiro models).

To narrow down these possibilities, it is crucial  to require   extra  UV information.  
For example, by demanding that the model has a chiral anomaly 
\cite{Albert:2023jtd,Ma:2023vgc},
  UV completions with  only scalar resonances are not allowed anymore.
Furthermore, the anomaly coefficient turned out to be bounded, providing new non-trivial constraints on the possible UV completions.

Here we would like to extend the analysis of \cite{Ma:2023vgc}  to  theories with $U(1)$-gravitational anomalies. 
For this purpose, we analyze here $2\to2$ amplitudes involving gravitons and the $U(1)$-Goldstone $\eta$, 
and study their dispersion relations based on analyticity, positivity and crossing.
Our main motivation is to obtain information on  possible 
 models of higher-spin  resonances
 that in the IR contain
a coupling of $\eta$ to two gravitons.

We first show that the $U(1)$-gravitational  anomaly coefficient $\kappa_g$ cannot be bounded when the graviton is an external (non-propagating) state.
We are then  forced to introduce  dynamical gravitons.  This brings 
the problem of dealing with the  $1/t$ pole caused by the exchange of gravitons
 in the  dispersion relations. 
 Fortunately, this problem has been overcome through the introduction of smearing techniques 
\cite{Caron-Huot:2021rmr, Caron-Huot:2022ugt, Caron-Huot:2022jli,Beadle:2024hqg} that have  already been  used in several physical systems  \cite{Henriksson:2022oeu, Hong:2023zgm, Albert:2024yap, Xu:2024iao, Haring:2024wyz} (for alternatives to smearing see for example \cite{Bellazzini:2019xts}).

Using these techniques,
we will show below  that 
dispersion relations in graviton-graviton scattering amplitudes  
lead to an upper bound    on the scale   at which   spin$\geq 4$ resonances 
must  appear, $\Lambda_{\text{caus}}$.  
The  bound will be rigorously and precisely derived, but  parametrically it  can be written as
 \be
  \Lambda_{\text{caus}}\sim \sqrt{\frac{M_PF_\pi}{\kappa_g}}\,,
\label{boundcausaI}
\ee
and it  is reminiscent of bounds found in time delay  analysis \cite{Camanho:2014apa,Serra:2022pzl}.
We will also study dispersion relations from  eta-graviton scattering amplitudes and show a similar bound holds for
 spin$\geq 2$ states coupled to $\eta h$.

For strongly-coupled  gauge theories  we will see  that 
the presence of glueballs makes the scale of quantum gravity smaller  than $\Lambda_{\text{caus}}$,
as long as  the number of fermions $N_F\ll N_c$, and the 'tHooft coupling ($g_{\rm YM}^2N_c$) is not very large. 
Therefore this analysis will not tell us much on the properties of gauge theories in this limit.

     The mass scale \eq{boundcausaI}  however will be especially  interesting in 5D holographic models where spin$>2$ states can be taken to be  infinitely heavy. These are models dual to   gauge theories in the large 'tHooft coupling.
          The requirement that $\Lambda_{\text{caus}}$  exceeds any other cutoff scale in the 5D models will impose non-trivial conditions on the 5D parameters.
 In particular, we will find  relations between 
 the couplings of  the spin-1 and spin-2 resonances and the Chern-Simons coefficient.

Causality bounds on large-$N_c$ theories   have been studied previously in
 \cite{Afkhami-Jeddi:2018apj, Kaplan:2019soo, Kaplan:2020ldi, Kaplan:2020tdz}. 
These analysis were based on scattering amplitudes of higher-spin resonances, assuming certain nonzero couplings among them. 
Our bounds however are simply based on the presence of the coupling of $\eta$ to two gravitons that is guaranteed  in all theories with a $U(1)$-gravitational anomaly.
Furthermore, 
we  extend the implications to theories with a large 'tHooft coupling, and 
 elaborate on the difference between the cutoff scale from causality  and other cutoff scales of the theory 
based on perturbativity such as the  quantum gravity scale.

The  paper is organized as follows. In Sec.~\ref{sec2} we introduce the 
$\eta hh$ coupling and its relation with the $U(1)$-gravitational anomaly.
In Sec.~\ref{sec3} we present all $2\to 2$ scattering amplitudes  involving the 
$U(1)$-gravitational anomaly coefficient $\kappa_g$.
 In Sec.~\ref{sec4} we briefly review the analyticity of scattering amplitudes and the derivation of dispersion relations. 
 In Sec. \ref{nobound}  we show that no bound can be obtained 
on  $\kappa_g$ when gravitons are  external frozen sources.
In Sec.  \ref{dynamgrav} we  consider  dynamical gravitons and 
show how to  derive  bounds using smearing techniques. 
 In Sec. \ref{pheno} we discuss the implications of the bounds  in axion models and   models of massive higher-spin resonances,
  and conclude in Sec. \ref{sec:conclude}.  
In three  Appendices we present technical details on the derivations of the bounds.
 In Appendix \ref{applargeJ} we  study the large $J$ limit of the smeared dispersion relations, 
and in Appendices \ref{2ndbound}  and \ref{elimit}
 we derive an  alternative bound on  $\kappa_g$ from graviton scattering
and  $\eta h^+ \to h^+h^+$ respectively,  making connection with the eikonal limit.

\section{The $\eta hh$ interaction and the chiral-gravitational anomaly}
\label{sec2}

We are interested in obtaining a constraint on the coupling of a pseudo-scalar,  $\eta$,
to two gravitons $h$. 
Using the spinor-helicity formalism \cite{Dixon:2015der}, we parametrize the three-point  on-shell coupling as  
\bea
 \label{etahh}
\begin{tikzpicture}[baseline=.1ex]
  \begin{feynman}[every dot={/tikz/fill=black!70}]
    \vertex[] (m1) at (-0.8, 0){};
    \vertex[dot][red] (m2) at (0.3, 0){};
    \vertex (a) at (-1,0) {$\eta$} ;
    \vertex (b) at ( 1.2,-1) {$h^{+}$};
    \vertex (d) at ( 1.2, 1) {\small{$h^{+}$}};
    \diagram* {
      (m2) -- [dashed] (m1),
      (b) -- [photon] (m2) -- [photon] (d)
    };
  \end{feynman}
  \end{tikzpicture}  \quad \quad \mathcal{M}_{\eta + +}= \frac{\kappa_g}{F_\pi M_P^2} [23]^4 \,,
\eea          
where $M_P$ is the Planck scale,\footnote{We define $1/M_P$  as the constant in the 3-point graviton interaction --see for example Eq.(44) of \cite{Baratella:2020dvw}.}
$F_\pi$ is the decay constant associated with the pseudo-scalar $\eta$,
and $\kappa_g$ a dimensionless coupling. Our normalization follows from the rule that there should be a $1/F_\pi$ for each Goldstone
and a $1/M_P$ for each graviton in the amplitude.
We also choose  the convention that all particles are incoming.

There are many interesting models where   the interaction \eq{etahh} is present. For example, in axion models
$\eta$ is identified with the axion, the Goldstone boson of a spontaneously broken  $U(1)$ global symmetry, 
and    $\kappa_g$ with the   $U(1)$-gravitational anomaly coefficient.
Also, \eq{etahh} is present  in  strongly-coupled theories with spontaneous breaking of the chiral symmetry,  with  $\kappa_g$ being  the chiral-gravitational anomaly coefficient. For example, in one-flavor QCD this is determined to be 
$\kappa_g ={N_c}/{192\pi^2}$.
Finally,  5D holographic models can also contain \eq{etahh} arising from  a Chern-Simons term.
All these types of models will be discussed in detail in Sec.~\ref{pheno}.
In all examples \eq{etahh} arises from the following term in the Lagrangian:
\be
\frac{i \kappa_g}{4M_P^2}   \ln\text{Det}[U]\,   
\epsilon^{\mu\nu\delta\gamma} {\cal R}_{\mu \nu \rho\sigma} {\cal R}_{\delta\gamma}^{\ \ \rho\sigma}  \,,
\label{qft}
\ee
where ${\cal R}_{ \mu \nu \rho\sigma}$ is the Riemann curvature tensor and 
 $\ln\text{Det}[U] = 2i \eta/F_\pi$.

To bootstrap $\kappa_g$ we could in principle proceed as in \cite{Ma:2023vgc} 
where a bound on the  $U(1)_A-SU(2)_V-SU(2)_V$ anomaly coefficient $\kappa$ was derived.
In this latter case,   one had the property that $\kappa$ also appeared as a local four-point interaction:
\be
{\cal M}_{\eta\pi_a\pi_b W^+_c}\propto \kappa\frac{[4124]}{F^3_\pi}f_{abc}\,,
\label{amp4ptc}
\ee
where $f_{abc}$ are the flavour structure constants and  $[ijkl]\equiv[i|p_jp_k|l]$.
Therefore it was possible to derive a  bound on $\kappa$  following the ordinary path of bounding  low-energy Wilson coefficients 
of $2\to 2$ processes  from analyticity and unitarity \cite{Albert:2023jtd,Ma:2023vgc}. 

Unfortunately, the situation is different for $\kappa_g$. As it is obvious from \eq{qft}, 
no  term with three Goldstones as \eq{amp4ptc} appears in this case.
The reason can be understood from simple dimensional analysis. 
A local 4-point interaction involving an odd number of $\eta$ should scale as
\bea 
{\cal M}_{\eta\eta\eta h^+}\sim\frac{[4124]^2}{M_PF^3_\pi M^2}
\quad,\quad
{\cal M}_{\eta h^+h^+h^+}\sim\frac{[23]^2[34]^2[42]^2}{M^3_PF_\pi M^2}
\quad,\quad
{\cal M}_{\eta h^-h^+h^+}\sim\frac{[34]^4\langle2342\rangle^2}{M^3_PF_\pi M^6}\,,
\label{amp4pt}
\eea
where,  contrary to \eq{amp4ptc}, we needed on dimensional grounds to introduce in all cases a new energy scale $M$. This means that
the amplitudes in \eq{amp4pt}  depend on the details  of the theory that determines $M$ and therefore are  not directly related with the anomaly. In other words, the coefficients in front of these operators cannot be   $\kappa_g$.

\section{Four-point  Amplitudes involving the $\eta h h$ interaction}
\label{sec3}

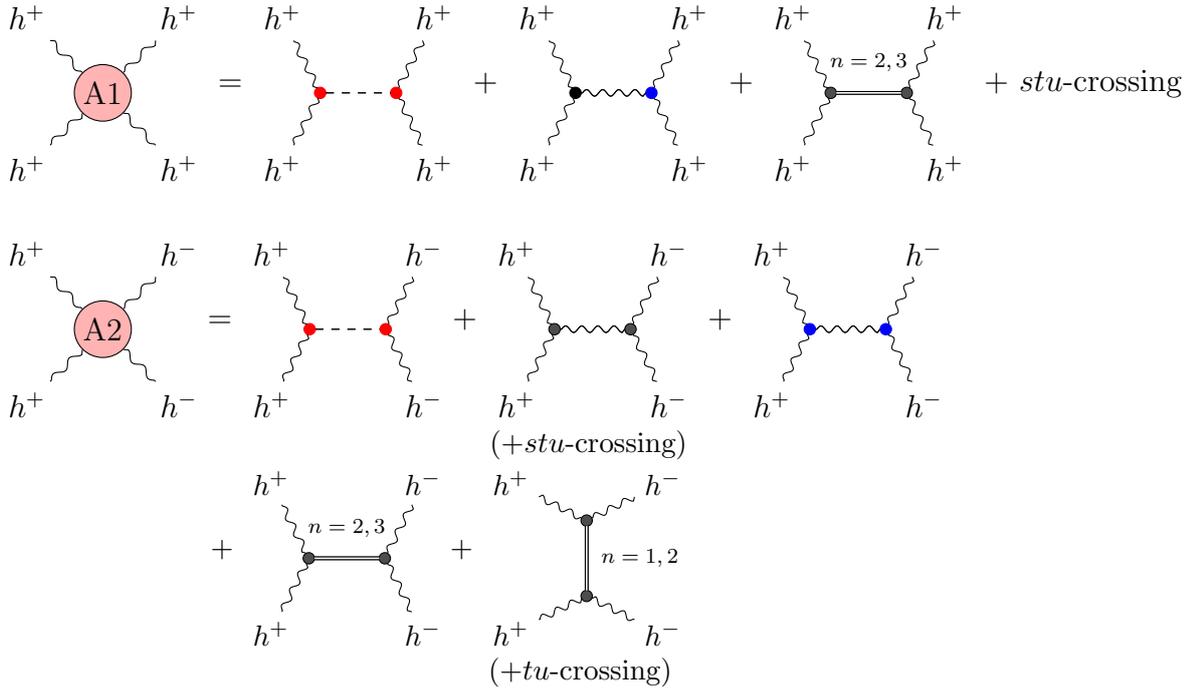
\begin{figure}[t]
    \begin{tikzpicture}[baseline=.1ex]
        \begin{feynman}[every blob={/tikz/fill=red!30}]
            \vertex[blob] (m) at (0, 0){};
            \vertex (e) at (0,0) {A1};
            \vertex (a) at (-1,-1) {$h^+$} ;
            \vertex (b) at ( 1,-1) {$h^+$};
            \vertex (c) at (-1, 1) {$h^+$};
            \vertex (d) at ( 1, 1) {$h^+$};
            \diagram* {
                (a) -- [photon] (m) -- [photon] (c),
                (b) -- [photon] (m) -- [photon] (d),
            };
        \end{feynman}
    \end{tikzpicture} = 
     \begin{tikzpicture}[baseline=.1ex]
        \begin{feynman}[every dot={/tikz/fill=black!70}]
            \vertex[dot][red] (m) at (0.5, 0){};
            \vertex[dot][red] (e) at (-0.5,0) {};
            \vertex (a) at (-1,-1) {$h^+$} ;
            \vertex (b) at ( 1,-1) {$h^+$};
            \vertex (c) at (-1, 1) {$h^+$};
            \vertex (d) at ( 1, 1) {$h^+$};
            \diagram* {
                (a) -- [photon] (e) --[dashed] (m) -- [photon] (b),
                (c) -- [photon] (e) --[dashed] (m) -- [photon] (d),
            };
        \end{feynman}
    \end{tikzpicture} +
         \begin{tikzpicture}[baseline=.1ex]
        \begin{feynman}[every dot={/tikz/fill=black!70}]
            \vertex[dot][blue] (m) at (0.5, 0){};
            \vertex[dot][black] (e) at (-0.5,0) {};
            \vertex (a) at (-1,-1) {$h^+$} ;
            \vertex (b) at ( 1,-1) {$h^+$};
            \vertex (c) at (-1, 1) {$h^+$};
            \vertex (d) at ( 1, 1) {$h^+$};
            \diagram* {
                (a) -- [photon] (e) --[photon] (m) -- [photon] (b),
                (c) -- [photon] (e) --[photon] (m) -- [photon] (d),
            };
        \end{feynman}
    \end{tikzpicture} +
  \begin{tikzpicture}[baseline=.1ex]
        \begin{feynman}[every dot={/tikz/fill=black!70}]
            \vertex[dot] (m) at (0.5, 0){};
                \vertex (n) at (0, 0.4){\scriptsize{$n=2,3$}};
            \vertex[dot] (e) at (-0.5,0) {};
            \vertex (a) at (-1,-1) {$h^+$} ;
            \vertex (b) at ( 1,-1) {$h^+$};
            \vertex (c) at (-1, 1) {$h^+$};
            \vertex (d) at ( 1, 1) {$h^+$};
            \diagram* {
                (a) -- [photon] (e) --[double] (m) -- [photon] (b),
                (c) -- [photon] (e) --[double] (m) -- [photon] (d),
            };
        \end{feynman}
    \end{tikzpicture} + $stu$-crossing \vspace{0.5cm}
\\
    \begin{tikzpicture}[baseline=.1ex]
        \begin{feynman}[every blob={/tikz/fill=red!30}]
            \vertex[blob] (m) at (0, 0){};
            \vertex (e) at (0,0) {A2};
            \vertex (a) at (-1,-1) {$h^+$} ;
            \vertex (b) at ( 1,-1) {$h^-$};
            \vertex (c) at (-1, 1) {$h^+$};
            \vertex (d) at ( 1, 1) {$h^-$};
            \diagram* {
                (a) -- [photon] (m) -- [photon] (c),
                (b) -- [photon] (m) -- [photon] (d),
            };
        \end{feynman}
    \end{tikzpicture}= 
     \begin{tikzpicture}[baseline=.1ex]
        \begin{feynman}[every dot={/tikz/fill=black!70}]
            \vertex[dot][red] (m) at (0.5, 0){};
            \vertex[dot][red] (e) at (-0.5,0) {};
            \vertex (a) at (-1,-1) {$h^+$} ;
            \vertex (b) at ( 1,-1) {$h^-$};
            \vertex (c) at (-1, 1) {$h^+$};
            \vertex (d) at ( 1, 1) {$h^-$};
            \diagram* {
                (a) -- [photon] (e) --[dashed] (m) -- [photon] (b),
                (c) -- [photon] (e) --[dashed] (m) -- [photon] (d),
            };
        \end{feynman}
    \end{tikzpicture}+
      \begin{tikzpicture}[baseline=.1ex]
        \begin{feynman}[every dot={/tikz/fill=black!70}]
            \vertex[dot] (m) at (0.5,0){};
            \vertex[dot] (e) at (-0.5,0) {};
            \vertex (a) at (-1,-1) {$h^+$} ;
            \vertex (b) at ( 1,-1) {$h^-$};
            \vertex (c) at (-1, 1) {$h^+$};
            \vertex (d) at ( 1, 1) {$h^-$};
            \diagram* {
                (a) -- [photon] (e) --[photon] (m) -- [photon] (b),
                (c) -- [photon] (e) --[photon] (m) -- [photon] (d),
            };
        \end{feynman}
    \end{tikzpicture} +     
     \begin{tikzpicture}[baseline=.1ex]
        \begin{feynman}[every dot={/tikz/fill=black!70}]
            \vertex[dot][blue] (m) at (0.5,0){};
            \vertex[dot][blue] (e) at (-0.5,0) {};
            \vertex (a) at (-1,-1) {$h^+$} ;
            \vertex (b) at ( 1,-1) {$h^-$};
            \vertex (c) at (-1, 1) {$h^+$};
            \vertex (d) at ( 1, 1) {$h^-$};
            \diagram* {
                (a) -- [photon] (e) --[photon] (m) -- [photon] (b),
                (c) -- [photon] (e) --[photon] (m) -- [photon] (d),
            };
        \end{feynman}
    \end{tikzpicture} \\
  ${}$ \hspace{6.2cm} \small{(+$stu$-crossing)}\\ 
   ${}$\hspace{2.8cm}+
  \begin{tikzpicture}[baseline=.1ex]
        \begin{feynman}[every dot={/tikz/fill=black!70}]
            \vertex[dot] (m) at (0.5, 0){};
            \vertex (n) at (0, 0.4){\scriptsize{$n=2,3$}};
            \vertex[dot] (e) at (-0.5,0) {};
            \vertex (a) at (-1,-1) {$h^+$} ;
            \vertex (b) at ( 1,-1) {$h^-$};
            \vertex (c) at (-1, 1) {$h^+$};
            \vertex (d) at ( 1, 1) {$h^-$};
            \diagram* {
                (a) -- [photon] (e) --[double] (m) -- [photon] (b),
                (c) -- [photon] (e) --[double] (m) -- [photon] (d),
            };
        \end{feynman}
    \end{tikzpicture}+
    \begin{tikzpicture}[baseline=.1ex]
        \begin{feynman}[every dot={/tikz/fill=black!70}]
            \vertex[dot] (m) at (0,0.5){};
            \vertex[dot] (e) at (0,-0.5) {};
            \vertex (n) at (0.7,0){\scriptsize{$n=1,2$}};
            \vertex (a) at (-1,-1) {$h^+$} ;
            \vertex (b) at ( 1,-1) {$h^-$};
            \vertex (c) at (-1, 1) {$h^+$};
            \vertex (d) at ( 1, 1) {$h^-$};
            \diagram* {
                (a) -- [photon] (e) --[double] (m) -- [photon] (c),
                (b) -- [photon] (e) --[double] (m) -- [photon] (d),
            };
        \end{feynman}
    \end{tikzpicture} \\ ${}$ \hspace{6.2cm} \small{(+$tu$-crossing)}
\caption{\it 
4-point amplitudes  involving $\kappa_g$  (red vertex) due to the exchange of the  pseudo-scalar $\eta$.  We also show the contributions from $\kappa_3$ (blue vertex) and ordinary gravity (black vertex).}
\label{fig:diagrams1}
\end{figure}
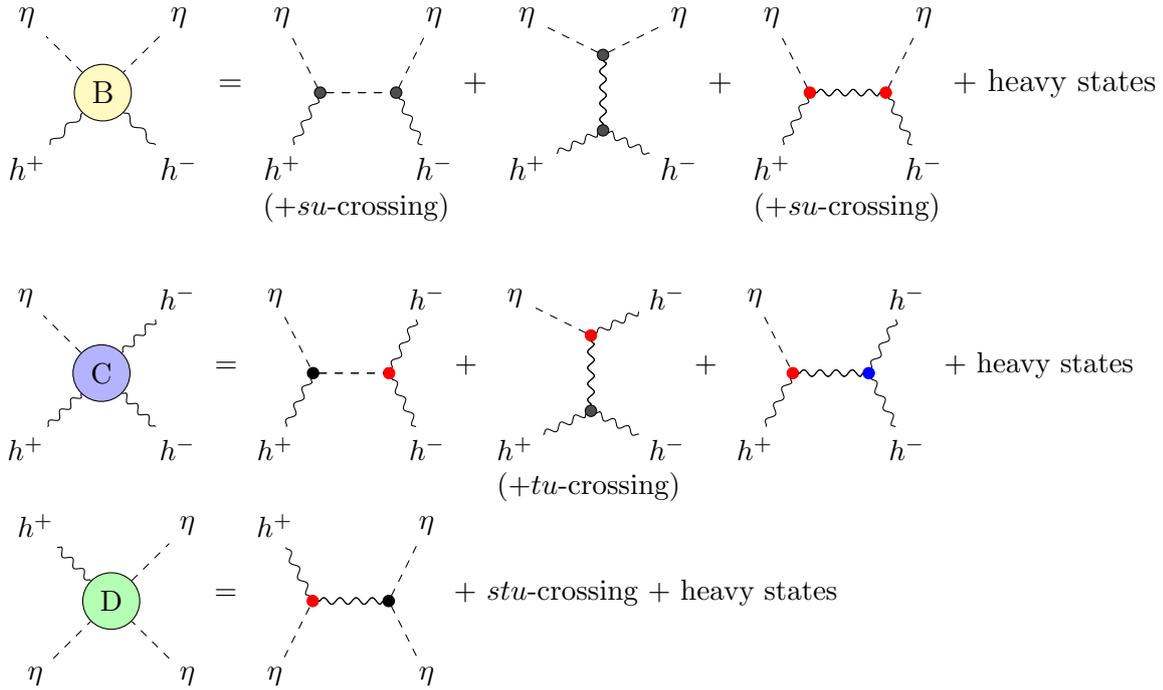
\begin{figure}[h]    
    
   \begin{tikzpicture}[baseline=.1ex]
        \begin{feynman}[every blob={/tikz/fill=yellow!30}]
            \vertex[blob] (m) at (0, 0){};
            \vertex (e) at (0,0) {B};
            \vertex (a) at (-1,1) {$\eta$} ;
            \vertex (b) at ( 1,1) {$\eta$};
            \vertex (c) at (-1, -1) {$h^+$};
            \vertex (d) at ( 1, -1) {$h^-$};
            \diagram* {
                (d) -- [photon] (m) -- [photon] (c),
                (b) -- [dashed] (m) -- [dashed] (a),
            };
        \end{feynman}
    \end{tikzpicture} = 
    \begin{tikzpicture}[baseline=.1ex]
        \begin{feynman}[every dot={/tikz/fill=black!70}]
            \vertex[dot] (m) at (-0.5,0){};
            \vertex[dot] (e) at (0.5,0) {};
            \vertex (a) at (-1,1) {$\eta$} ;
            \vertex (b) at ( 1,1) {$\eta$};
            \vertex (c) at (-1, -1) {$h^+$};
            \vertex (d) at ( 1, -1) {$h^-$};
            \diagram* {
                (a) -- [dashed] (m) --[dashed] (e) -- [dashed] (b),
                (c) -- [photon] (m) --[dashed] (e) -- [photon] (d),
            };
        \end{feynman}
    \end{tikzpicture}+    
    \begin{tikzpicture}[baseline=.1ex]
        \begin{feynman}[every dot={/tikz/fill=black!70}]
            \vertex[dot] (m) at (0,0.5){};
            \vertex[dot] (e) at (0,-0.5) {};
            \vertex (a) at (-1,-1) {$h^+$} ;
            \vertex (b) at ( 1,-1) {$h^-$};
            \vertex (c) at (-1, 1) {$\eta$};
            \vertex (d) at ( 1, 1) {$\eta$};
            \diagram* {
                (a) -- [photon] (e) --[photon] (m) -- [dashed] (c),
                (b) -- [photon] (e) --[photon] (m) -- [dashed] (d),
            };
        \end{feynman}
    \end{tikzpicture}+    \begin{tikzpicture}[baseline=.1ex]
        \begin{feynman}[every dot={/tikz/fill=black!70}]
            \vertex[dot][red] (m) at (-0.5,0){};
            \vertex[dot][red] (e) at (0.5,0) {};
            \vertex (a) at (-1,1) {$\eta$} ;
            \vertex (b) at ( 1,1) {$\eta$};
            \vertex (c) at (-1, -1) {$h^+$};
            \vertex (d) at ( 1, -1) {$h^-$};
            \diagram* {
                (a) -- [dashed] (m) --[photon] (e) -- [dashed] (b),
                (c) -- [photon] (m) --[photon] (e) -- [photon] (d),
            };
        \end{feynman}
    \end{tikzpicture}+ heavy states \\   ${}$ \hspace{3.35cm}\small{(+$su$-crossing)}\hspace{4cm}\small{(+$su$-crossing)}   \\ \vspace{0.3cm}

    \begin{tikzpicture}[baseline=.1ex]
        \begin{feynman}[every blob={/tikz/fill=blue!30}]
            \vertex[blob] (m) at (0, 0){};
            \vertex (e) at (0,0) {C};
            \vertex (a) at (-1,-1) {$h^+$} ;
            \vertex (b) at ( 1,-1) {$h^-$};
            \vertex (c) at (-1, 1) {$\eta$};
            \vertex (d) at ( 1, 1) {$h^-$};
            \diagram* {
                (a) -- [photon] (m) -- [dashed] (c),
                (b) -- [photon] (m) -- [photon] (d),
            };
        \end{feynman}
    \end{tikzpicture} = 
    \begin{tikzpicture}[baseline=.1ex]
        \begin{feynman}[every dot={/tikz/fill=black!70}]
            \vertex[dot][red] (m) at (0.5, 0){};
            \vertex[dot][black] (e) at (-0.5,0) {};
            \vertex (a) at (-1,-1) {$h^+$} ;
            \vertex (b) at ( 1,-1) {$h^-$};
            \vertex (c) at (-1, 1) {$\eta$};
            \vertex (d) at ( 1, 1) {$h^-$};
            \diagram* {
                (a) -- [photon] (e) --[dashed] (m) -- [photon] (b),
                (c) -- [dashed] (e) --[dashed] (m) -- [photon] (d),
            };
        \end{feynman}
    \end{tikzpicture}+    
    \begin{tikzpicture}[baseline=.1ex]
        \begin{feynman}[every dot={/tikz/fill=black!70}]
            \vertex[dot][red] (m) at (0,0.5){};
            \vertex[dot] (e) at (0,-0.5) {};
            \vertex (a) at (-1,-1) {$h^+$} ;
            \vertex (b) at ( 1,-1) {$h^-$};
            \vertex (c) at (-1, 1) {$\eta$};
            \vertex (d) at ( 1, 1) {$h^-$};
            \diagram* {
                (a) -- [photon] (e) --[photon] (m) -- [dashed] (c),
                (b) -- [photon] (e) --[photon] (m) -- [photon] (d),
            };
        \end{feynman}
    \end{tikzpicture}+     \begin{tikzpicture}[baseline=.1ex]
        \begin{feynman}[every dot={/tikz/fill=black!70}]
            \vertex[dot][blue] (m) at (0.5, 0){};
            \vertex[dot][red] (e) at (-0.5,0) {};
            \vertex (a) at (-1,-1) {$h^+$} ;
            \vertex (b) at ( 1,-1) {$h^-$};
            \vertex (c) at (-1, 1) {$\eta$};
            \vertex (d) at ( 1, 1) {$h^-$};
            \diagram* {
                (a) -- [photon] (e) --[photon] (m) -- [photon] (b),
                (c) -- [dashed] (e) --[photon] (m) -- [photon] (d),
            };
        \end{feynman}
    \end{tikzpicture} + heavy states \\   ${}$ \hspace{6.31cm} \small{(+$tu$-crossing)}\\ \vspace{0.3cm}
    \begin{tikzpicture}[baseline=.1ex]
        \begin{feynman}[every blob={/tikz/fill=green!30}]
            \vertex[blob] (m) at (0, 0){};
            \vertex (e) at (0,0) {D};
            \vertex (a) at (-1,-1) {$\eta$} ;
            \vertex (b) at ( 1,-1) {$\eta$};
            \vertex (c) at (-1, 1) {$h^+$};
            \vertex (d) at ( 1, 1) {$\eta$};
            \diagram* {
                (a) -- [dashed] (m) -- [photon] (c),
                (b) -- [dashed] (m) -- [dashed] (d),
            };
        \end{feynman}
    \end{tikzpicture} =  \begin{tikzpicture}[baseline=.1ex]
        \begin{feynman}[every dot={/tikz/fill=black!70}]
            \vertex[dot][black] (m) at (0.5, 0){};
            \vertex[dot][red] (e) at (-0.5,0) {};
            \vertex (a) at (-1,-1) {$\eta$} ;
            \vertex (b) at ( 1,-1) {$\eta$};
            \vertex (c) at (-1, 1) {$h^+$};
            \vertex (d) at ( 1, 1) {$\eta$};
            \diagram* {
                (a) -- [dashed] (e) --[photon] (m) -- [dashed] (b),
                (c) -- [photon] (e) --[photon] (m) -- [dashed] (d),
            };
        \end{feynman}
    \end{tikzpicture}  + $stu$-crossing + heavy states
 
\caption{\it 
4-point amplitudes  involving  $\kappa_g$ (red vertex) due to the exchange of the 
graviton $h$ (in C together with  the  $\eta$ exchange).
We also show the contributions from $\kappa_3$ (blue vertex) and ordinary gravity (black vertex).}
\label{fig:diagrams2}
\end{figure}

As we have seen in the previous section, there are no 4-point amplitudes that  contain $\kappa_g$
as a  contact interaction  at  low energies.
Therefore  $\kappa_g$ can only enter in  $2\to 2$ amplitudes  via the  3-point $\eta hh$ coupling \eq{etahh},
necessarily requiring also  the mediation of either a $\eta$ or a  graviton. We will be interested in amplitudes only involving
massless external states as this simplifies the analysis of bootstrapping $\kappa_g$.
 This means amplitudes of  Goldstones and gravitons.

In Fig.~\ref{fig:diagrams1} we show all possible 4-point amplitudes where $\kappa_g$ contributes exclusively via the exchange of the pseudo-scalar $\eta$ (no graviton exchange required).
The amplitude  A1 has an internal propagator of $\eta$ in the $s$, $t$ and $u$ channel,
while in the amplitude A2 the internal propagator of $\eta$ appears only in the $s$ channel,
giving in both cases a contribution  proportional to $\kappa_g^2$.
These are amplitudes which in principle could allow us to  bootstrap $\kappa_g$ 
in the decoupling limit of gravity $M_P\to \infty$ (the graviton acting as an external non-dynamical source).\footnote{Notice that the graviton can also appear as an internal line in A2 via ordinary gravity (black vertex), as seen in Fig.~\ref{fig:diagrams1}, but we can use dispersion relations that avoid this contribution.}
Unfortunately, as we will show in the next section, we will not be able to bound $\kappa_g$ in this limit.

In Fig.~\ref{fig:diagrams2}  we show the 4-point amplitudes where  $\kappa_g$ contributions     comes from diagrams involving the exchange of a dynamical graviton. 
In particular,  the amplitude B  (${\cal M}_{\eta + \eta -}$) contains
 a $\kappa^2_g$ contribution arising  from the exchange of  a graviton 
  in the $s$  and $u$ channels,
   while  the amplitude D    (${\cal M}_{\eta\eta\eta+}$)
   contains a  $\kappa_g$ contribution   from exchanging a graviton in the $s$ and $u$ channels.
 In the amplitude C  (${\cal M}_{\eta +--}$)  a
   $\kappa_g$ contribution
comes from the pseudo-scalar $\eta$ exchange in the $s$ channel and the graviton in the $t$ and $u$ channels. Note  that in this latter case the graviton needs to be present by gauge  invariance and cannot be disentangled from the $\eta$ exchange.
In principle, also the 4-point amplitude  ${\cal M}_{\eta +++}$ could  have  contributions proportional to $\kappa_g^2$
from $\eta$ and graviton exchange  but these contributions turn out to be  zero.

In Fig.~\ref{fig:diagrams1} and Fig.~\ref{fig:diagrams2} we have also introduced  for completeness the 3-point  interaction of  equal-helicity  gravitons  (blue vertices) defined as
\be
 \mathcal{M}_{+++} = \frac{\kappa_3}{M_P^3} [12]^2[23]^2[13]^2\,,
\label{k3}
\ee
  where  we note that $\kappa_3$ has dimension $1/E^2$.
This corresponds  to the Riemann curvature ${\cal R}^3$ operator.
The interaction \eq{k3} will  be turned off 
in Section \ref{nobound}, where  gravitons are treated as non-dynamical probes.
In Section \ref{dynamgrav} on the other hand, when taking the graviton as a dynamical field, we will  have to take  this  interaction into account.

\section{Sum rules  from dispersion relations}
\label{sec4}

\begin{figure}[t]
  \begin{center}
\begin{tikzpicture}
\draw[decoration={markings, mark=at position 0.625 with {\arrow{>}}},postaction={decorate}, color=gray] (3,0) ellipse (2.8 and 1.8) {};
\draw[decoration={markings, mark=at position 0.625 with {\arrow{>}}},postaction={decorate}, color=blue] (3,0) circle (0.3) {};
\node[color=gray] at (1.8,-1.3) {\footnotesize{$C_\infty$}};
\node[color=blue] at (2.72,0.5) {\footnotesize{$C_0$}};
\draw[black] (5,2) -- (5,1.8) node[right] {$s$};
\draw[black] (5,1.8) -- (5,1.6) {};
\draw[black] (5,1.6) -- (5.4,1.6) {};
\draw[black, thick,->] (0,0) -- (6,0);
\draw[black, thick,->] (3,-2) -- (3,2);
\draw (3,0) node[crossr] {};
\draw (3.9,0) node[crossp] {};
\draw (4.4,0) node[crossp] {};
\draw (4.9,0) node[crossp] {};
\draw (5.4,0) node[crossp] {};
\draw (0.6,0) node[crossp] {};
\draw (1.1,0) node[crossp] {};
\draw (1.6,0) node[crossp] {};
\draw (2.1,0) node[crossp] {};
\draw[color=blue] (3.8,0.2)--(4.65,0.2) {};
\draw[color=blue,<-] (4.65,0.2)--(5.7,0.2) {};
\draw[color=blue] (3.8,-0.2)--(5.7,-0.2) {};
\draw[color=blue] (3.8,0.2) arc (90:270:0.2) node[midway,below=0.18cm,align=center] {\textcolor{black}{$\qquad M_s^2  $}};
\draw[color=blue] (0.4,0.2)--(2.2,0.2) {};
\draw[color=blue,->] (0.4,-0.2)--(1.3,-0.2) {};
\draw[color=blue] (1.3,-0.2)--(2.2,-0.2) {};
\draw[color=blue] (2.2,0.2) arc (90:-90:0.2) node[midway,below=0.18cm,align=center] {\textcolor{black}{$-M_u^2\hspace{-0.1cm}-t  \qquad\  $}};
\end{tikzpicture} 
\label{fig:utchannel}
  \end{center}
  \caption{\it Analytic structure of a generic ${\cal M}(s,u)$ amplitude at fixed $t<0$. The positive real $s$-axis contains $s$-channel poles, while the negative real $s$-axis has $u$-channel poles. We also show the contours used to obtain dispersion relations.}
    \label{fig:sumrules}
\end{figure}
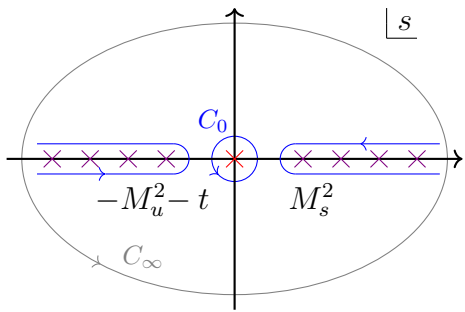

Let us start by reviewing  how requiring analyticity, crossing and unitarity of scattering amplitudes
leads to constrains on  low-energy parameters. We consider amplitudes mediated by a tree-level exchange of states.

For $2\to 2$ processes at real $t<0$ fixed, causality tells us  that the corresponding amplitudes must be analytic in the whole complex 
$s$-plane except on the real $s$-axis. For amplitudes mediated by a tree-level exchange of particles, only simple poles can appear on  the real $s$-axis, as illustrated in Fig.~\ref{fig:sumrules}.  
Whenever  an integral along the contour $C_\infty$ as in Fig.~\ref{fig:sumrules}  vanishes,
one can use Cauchy's theorem to relate the low-energy integral (along the contour $C_0$) with the high-energy integral contouring the non-analytic contribution of the amplitude. 

 We will assume that  our amplitudes  at $t$-fixed have the following high-energy  behaviour
\bea 
\lim_{|s|\rightarrow \infty} \frac{\mathcal{M}(s,u)}{s^k}\rightarrow 0 \ \ \text{for all}\ \  k\geq k_\text{min}\,,
\label{eq:UV}
\eea
where the smallest value of $k$, defined as $k_\text{min}$, will be specified later for each particular case.

Dispersion relations  can be derived in the following way. 
We start with the integral of ${\cal M}(s,-s-t)/s^{k+1}$ along the contour $C_\infty$ of Fig.~\ref{fig:sumrules} which
 vanishes for $k\geq{k_{\rm min}}$ due to \eq{eq:UV}. 
 Because of the amplitude's analyticity, we can deform $C_\infty$ into the blue contour in Fig.~\ref{fig:sumrules} and obtain
\be
-\frac{1}{2 i}\oint_{\rm C_0} ds'\ \frac{{\cal M} (s',-t-s')}{s^{\prime k+1}}=
\int_{M_s^2}^\infty  ds'\ \frac{{\rm Im} {\cal M}(s',-t-s')}{s^{\prime k+1}}
+(-1)^k\int_{M_u^2}^\infty  ds'\ \frac{{\rm Im} {\cal M}(-t-s',s')}{(s'+t)^{k+1}}\,,
\label{DR}
\ee
where $M_{s,u}$ represents the mass of the lightest state in the $s$-channel ($u$-channel).  
 The partial-wave expansion for the imaginary part of the amplitude  is given by
\bea 
\text{Im} {\cal M}_{\lambda_1,\lambda_2,\lambda_3,\lambda_4}(s,\theta_s)&=&\sum_J(2J+1)\,\rho_J(s)\,d^J_{\lambda_1-\lambda_2,\lambda_4-\lambda_3}\left(\cos\theta_s\right)\,,
\nonumber\\
(2J+1)\rho_J(s)&=&\pi \sum_i g^2_i \,m_i^2\,\delta(s-m_i^2)\delta_{JJ_i}\,,
\label{eq:imM}
\eea
where $\lambda_i$ are the helicities of the external particles, $g^2_i$ is the residue of the state $i$ exchanged  in the $s$ channel, $ \cos\theta_s = 1+2t/s $ and $d^J_{m,m'}$ are the Wigner $d$-functions.

For the UV contribution to the dispersion relations it is convenient to define the high energy average
\be
\big\langle (...) \big\rangle \equiv\frac{1}{\pi}\sum_J (2J+1)  \int_{M_{s,u}^2}^\infty  \frac{dm^2}{m^2}\rho_J(m^2) (...)\,.
\ee
Using this notation, \eq{DR} can be written as
\be
-\frac{1}{2 i}\oint_{\rm C_0} ds'\ \frac{{\cal M} (s',-t-s')}{s^{\prime k+1}} = \left\langle \frac{d^J_{\lambda_1-\lambda_2,\lambda_4-\lambda_3}(\cos\theta_s)}{m^{2k}} \right \rangle +(-1)^k \left\langle \frac{d^J_{\lambda_1-\lambda_4,\lambda_2-\lambda_3}(\cos\theta_s) m^2}{(m^{2} +t)^{k+1}} \right \rangle \,,
 \label{eq:DRsecond}
\ee
where the first (second) term of the RHS is the $s$-channel ($u$-channel) contribution.

\section{Attemping to   bound  $\kappa_g$ without dynamical gravitons}
\label{nobound}

Let us first attempt  to study the four-point graviton amplitudes A1 and A2  of Fig.~\ref{fig:diagrams1} where  
the graviton can be taken as an external "probing" source, and, potentially,  one could obtain a  bound on $\kappa_g$ 
independent  of $M_P$.  We will however see that following this strategy
{\it no bound can be constructed}.

 \begin{table}[t]
\begin{center}
\renewcommand{\arraystretch}{1.6}
\begin{NiceTabular}[corners,hvlines]{|c||c|}
  $n$ & $J^P$ 
      \\
 \hline
1 &  $J_\text{odd}^+$ ($f_{J_\text{odd}}$)  \\
2 &  $J_\text{even}^+$  ($f_{J_\text{even}}$)  \\
3 & $J_\text{even}^- $  ($\eta_J$)    \\
                  \hline
\end{NiceTabular}
\end{center}
\caption{\it List of  states labeled by $n$ that enter in  4-graviton amplitudes classified in terms of their spin and parity $J^P$. We have defined $J_{\rm even}\equiv0,2,...$ and $J_{\rm odd}\equiv1,3,...$.  
We also provide the names as they are generally referred to in QCD \cite{ParticleDataGroup:2022pth}. }
\label{tab:states}
\end{table}

The two amplitudes to consider are (A1) ${\cal M}_{++++}$  associated to the inelastic process
$h^+h^+\to h^-h^-$ and  (A2) ${\cal M}_{++--}$ 
associated to the elastic process
$h^+h^+\to h^+h^+$.
These amplitudes are mediated by the exchange of states that can be classified according to 
their  parity ($P$) and spin $J$  \cite{Jacob:1959at}. 
The states mediating A1 and A2 are classified in Table~\ref{tab:states} and in the text we will label them with $n=1,2,3$.

It is easy to derive which states enter in a given process.
In particular, one can determine that  $h^+h^+$   can only couple to spin-even  states
    due to identical particle permutation symmetry,\footnote{Under the interchange of two identical gravitons the wave-function changes as $(-1)^J$    where $J$ is the total angular momentum.  Being bosons, we must require $J$ to be even.} i.e.,  states $n=2$ and $n=3$. 
On the other hand, $h^+h^-$ is a system  of parity even, so it can only couple to states $n=1$ and $n=2$. 
This completely fixes the particle exchange in A1 and A2 as illustrated in Fig.~\ref{fig:diagrams1}.

At energies below the mass  of the lightest massive state ($M$), we can expand the amplitudes in powers of $s/M^2$ and $t/M^2 \to 0$ 
and obtain (after stripping the little group helicity phase, which does not affect the analyticity)
\bea
{\cal M}_{++++} &\to& -\frac{3\kappa_g^2}{M_P^4F_\pi^2}stu+a_{4,0}(s^4+t^4+u^4)+\dots\,, \label{eq:pppp}\\
{\cal M}_{++--} &\to&  s^4\left( -\frac{1}{M_{P}^2 stu}+\frac{\kappa_g^2}{F_\pi^2 M_{P}^4s} -b_{0,0}+\dots\right) \label{eq:ppmm}\,,
\eea
where $a_{4,0},b_{0,0}$  are low-energy constants, also referred to as Wilson coefficients.
The first term of  ${\cal M}_{++--}$ arises from the  graviton exchange (which would require the graviton to be dynamical), but, 
as we will see,  it is not picked up by the dispersion relations.

Armed with the above, we can derive  sum rules for the anomaly coefficient $\kappa_g$ 
from the A1 and A2 processes and attempt to bound it.

\subsection{Inelastic dispersion relations: $h^+ h^+ \to h^- h^-$}
First let us consider process A1. Since this scattering amplitude is $stu$-symmetric, there is only one set of independent sum rules. We look at the $t$-fixed dispersion relation in the complex $s$-plane. 

We  take $k_\text{min}=2$ in \eq{eq:UV}.
This can be argued from  the Regge theory which states that an amplitude
 at $|s|\to\infty$  and fixed $t$  goes as ${\cal M}\to s^{\alpha(0)}$  where 
$\alpha(0)$  is  the largest value that a  Regge trajectory   can take at $t=0$. 
The Regge trajectory corresponds to an analytic continuation of the spin of the states 
 exchanged in the $t$ channel:  $J=\alpha(t)$.
If the lightest state of this  trajectory has spin $J$, 
it follows from the positivity of the slope of the Regge trajectories that $\alpha(0)$  must necessarily 
be less than or equal to $J$, and therefore  $k_\text{min}\leq J$.
 Further details can be found in \cite{Gribov:2003nw,Albert:2023jtd}. 
  For  the process $h^+ h^+ \to h^- h^-$,  where   $J$ even  state   ($n=2,3$) are exchanged in the $t$ channel,
we conservatively assume  that the lightest states has $J\leq 2$.
 Therefore all trajectories satisfy $\alpha(0) <2$ and we can take   $k_\text{min}=2$.\footnote{Even if we took  $k_\text{min}=1$, we would get the same sum rules.} 
Inserting Eq. \eqref{eq:pppp} in Eq. \eqref{eq:DRsecond}, we get
\begin{alignat}{3}
k=2 &   &\qquad\frac{3\kappa_g^2}{M_P^4F_\pi^2}t + 6a_{4,0}t^2+\dots+\Big\langle \frac{\,d^J_{0,0}(1+2t/m^2)}{(m^2)^2} \Big\rangle + \Big\langle \frac{m^2\,d^J_{0,0}(1+2t/m^2)}{(m^2+t)^3} \Big\rangle &{}={}& & 0\,, \nonumber\\
k=3 &   &4a_{4,0}t+\dots+\Big\langle \frac{m^2\,d^J_{0,0}(1+2t/m^2)}{(m^2)^4} \Big\rangle - \Big\langle \frac{m^2\,d^J_{0,0}(1+2t/m^2)}{(m^2+t)^4} \Big\rangle &{}={}& & 0\,, \\
\vdots \ \ &&&& \nonumber
\end{alignat}
Expanding the $k=2$ dispersion relation at small $t$, we can extract the following sum rules
\bea
\bigg\langle  \frac{1}{m^4}\bigg \rangle^{g_{++}^2}_{2-3} = 0\,
\qquad\text{and}\qquad
-\frac{3\kappa_g^2}{M_P^4F_\pi^2}= \bigg\langle  \frac{2\mathcal{J}^2-3}{m^6}\bigg \rangle^{g_{++}^2}_{2-3},
\label{eq:DR2}
\eea
where the  superindex $g_{++}^2$ refers to the  coupling that enters in the residues, i.e.,  
 $g^2_i\propto  g_{++,i}^2$, 
 and the  subindex $2-3$ refers to the type of exchanged state and the sign of the residue:\footnote{ These signs are fixed by parity  and time reversal  conservation, along the lines  of Appendix A of \cite{Ma:2023vgc}. 
 One finds  ${\cal M}(R\to h^-h^-)=P_R \times {\cal M}^*(h^+h^+\to R)$ where $R$ is the resonance exchanged and $P_R$ its parity. This implies that   ${\cal M}(h^+h^+\to R)\times {\cal M}(R\to h^-h^-)=P_R|\mathcal{M}(h^+h^+\to R)|^2$ so its sign is determined by $P_R$.}
 \bea
 n=2\ {\rm states\qquad with } && g^2_i=  g_{++,i}^2\geq 0\,,\nonumber\\
 n=3\ {\rm  states \qquad with} && g^2_i=  -g_{++,i}^2\leq 0\,.
\eea
\eq{eq:DR2}  are the only sum rules up to order $1/m^6$. 

\subsection{Elastic dispersion relations: $h^+ h^+ \to h^+ h^+$}
The elastic process A2  is $tu$-symmetric, and therefore one can obtain two linearly independent sets of dispersion relations: the fixed $t$ (or $u$) dispersion relation in the complex $s$-plane and the $s$ fixed dispersion relation in the complex $t$ (or $u$) plane. 
The latter dispersion relations  give however  sum rules starting at  $O(1/m^{10})$, being irrelevant for $\kappa_g$.

Let us therefore focus on the amplitude A2 in the complex $s$-plane at fixed $t$.
To avoid the graviton exchange  (see Fig.~\ref{fig:diagrams1}),  we  take  $k_\text{min}=3$.
Inserting Eq. \eqref{eq:ppmm} in Eq. \eqref{DR}, we get the following relations
\bea
k=3 && \quad \frac{\kappa_g^2}{M_P^4 F_\pi^2} 
+ \Big\langle \frac{m^2\,d^J_{0,0}(1+2t/m^2_i)}{(m^2)^4} \Big\rangle^{g_{++}^2} - \Big\langle \frac{m^2\,d^J_{4,4}(1+2t/m^2)}{(m^2+t)^4} \Big\rangle^{g_{+-}^2} = 0\,, 
\\
\vdots \ \ && \nonumber
\eea
for which a small $t$ expansion gives the following sum rule
\bea
\label{eq:DR1}
\frac{\kappa_g^2}{M_P^4 F_\pi^2} = \bigg\langle \frac{1}{m^6}\bigg\rangle^{g_{+-}^2}_{1+2} -\bigg\langle \frac{1}{m^6}\bigg\rangle^{g_{++}^2}_{2+3} \,.
\eea
Other sum rules can be obtained from $k>3$, but these are of $O(1/m^8)$ or higher.

\subsection{Regge improved dispersion relations}

We can also  improve the high-energy behaviour of an amplitude  
 by taking superpositions of initial and final states such that   the contributions from the leading Regge trajectory is absent \cite{Albert:2023jtd}.
Consider the combination ${\cal M}_{++++} - {\cal M}_{++--}$ in the complex 
$t$-plane  at fixed  $s$. 
Now the high-energy behaviour is dictated by the Regge trajectory in the $s$-channel 
that does not contain any $n=2$ state since they cancel out in the combination ${\cal M}_{++++} - {\cal M}_{++--}$ (see previous sections).
The leading Regge trajectory consists  only of $n=3$ states, and the lightest one corresponds
to the Goldstones with $J=0$.  Therefore Regge theory tells us that we can take $k_{\rm min}=0$.

The $k=0$  dispersion relation for  ${\cal M}_{++++} - {\cal M}_{++--}$
 in the complex $t$ (or $u$) plane  at fixed  $s$ leads to 
\bea
\frac{\kappa_g^2}{M_P^4F_\pi^2}s^3- (b_{0,0}+ 2 a_{4,0})s^4+\dots = \hspace{8cm}\nonumber\\
\Bigg\langle d^J_{0,0}(1+2s/m^2)\bigg(1+\frac{m^2}{m^2+s}\bigg)\Bigg\rangle^{g_{++}^2}- \Bigg\langle d^J_{4,-4}(1+2s/m^2)\bigg(1+\frac{m^2}{m^2+s}\bigg)\Bigg\rangle^{g_{+-}^2}\,,
\eea
from which one gets the following sum rules and null constraints up to $O(1/m^6)$:
\bea
\label{eq:DR3}
\bigg\langle\ 1\ \bigg\rangle^{g_{++}^2}_{2-3} &=& 0\,,\nonumber\\
\bigg\langle\frac{2\mathcal{J}^2-1}{m^2}\bigg\rangle^{g_{++}^2}_{2-3} &=& 0\,,\nonumber\\
\bigg\langle\frac{\mathcal{J}^4/2-\mathcal{J}^2+1}{m^4}\bigg\rangle^{g_{++}^2}_{2-3}&=& 0\,,\nonumber\\
\bigg\langle\frac{\mathcal{J}^6/18-\mathcal{J}^4/4+\mathcal{J}^2-1}{m^6}\bigg\rangle^{g_{++}^2}_{2-3} &=& \frac{\kappa_g^2}{M_P^4 F_\pi^2} \,.
\eea

\subsection{No bound on $\kappa_g$}
\eq{eq:DR2}, \eq{eq:DR1}  and \eq {eq:DR3} give the complete set of  sum rules  up to $O(1/m^6)$. 
Unfortunately, they are not enough to constrain $\kappa_g$. 
The reason is the following. 
The anomaly coefficient, as it appears in the dispersion relation, scales as
\be
\frac{\kappa_g^2}{M_P^4 F^2_\pi} \propto \frac{1}{M_P^4m^6}\,.
\ee
 Nevertheless,    the  other Wilson coefficients scale as
 \be
 a_{4,0},b_{0,0},...\propto  \frac{1}{M^4_Pm^p}\ \ (p>6)\,.
\ee
  Therefore we cannot get an upper bound on ${\kappa_g^2}/{M_P^4 F^2_\pi}$ as a function of the Wilson coefficients as the latter are always smaller  in the large mass limit $m\to\infty$.

Neither a lower bound on $\kappa_g$ can  be obtained.
From the three sum rules that determine $\kappa_g$,  \eq{eq:DR2}, \eq{eq:DR1} and \eq{eq:DR3},
we see that the  contributions from the different $n=1,2,3$ states do not appear with the same sign.
Therefore it is always possible to tune the couplings of these states to have  ${\kappa_g^2}/{M_P^4 F^2_\pi}=0$. 
 This tuning however does not have to affect 
the  other Wilson coefficients $a_{4,0},b_{0,0},...$ since these are determined by different sum rules with different dependence  on the mass spectrum (none at $O(1/m^6$)).  

Therefore we conclude that no bound can be obtained for $\kappa_g$ from the above dispersion relations.

\section{$M_P$ dependent bound on $\kappa_g$ (dynamical gravitons)}
\label{dynamgrav}

The reason why we could not bound $\kappa_g$ in the previous section was  that we did  not find 
any  positive-definite  low-energy constant  up to order $1/m^6$. 
In this section we will see that 
by  allowing the graviton to be an exchanged particle, we will be able to access a sum rule at $O(1/m^4)$, involving  $M_P^2$,  which can be constructed to be positive-definite.  The price to pay will be that due to the graviton pole $1/t$, we will not be able to take the $t\to0$ limit but instead we will have to integrate the sum rule over $t$, weighted by some "smearing" function \cite{Caron-Huot:2021rmr}.

\subsection{Graviton scattering}
\label{gravgrav}

Let us begin by considering the elastic process $h^+ h^+ \rightarrow h^+h^+$  (Amplitude A2 in Fig.~\ref{fig:diagrams1}). Allowing for the exchange of gravitons, the low-energy amplitude becomes
\bea
{\cal M}_{++--} &\to& s^4\left(-\frac{1}{M_{P}^2 stu}+\frac{\kappa_g^2}{F_\pi^2 M_{P}^4s} -\frac{\kappa_3^2}{ M_{P}^6}\frac{tu}{s}-b_{0,0}+\dots\right) \,,
\label{mpm}
\eea
where we also introduce  the coupling $\kappa_3$ for generality since the gravitons are now dynamical. 
The relevant dispersion relations starting at $k_\text{min}=2$ are
\bea
k=2\hspace{3.3cm}  \frac{-1}{M_P^2\ t} &=& \Big\langle \frac{d^J_{0,0}(1+2t/m^2)}{(m^2)^2} \Big\rangle^{g_{++}^2} + \Big\langle \frac{m^2\,d^J_{4,4}(1+2t/m^2)}{(m^2+t)^3} \Big\rangle^{g^2_{+-}}\,,\label{drmp}\\
k=3\hspace{1.5cm} \frac{-\kappa_g^2}{M_P^4 F_\pi^2} -\frac{\kappa_3^2}{ M_{P}^6}t^2 &=& \Big\langle \frac{d^J_{0,0}(1+2t/m^2)}{(m^2)^3} \Big\rangle^{g_{++}^2} - \Big\langle \frac{m^2\,d^J_{4,4}(1+2t/m^2)}{(m^2+t)^4} \Big\rangle^{g^2_{+-}}\,, \label{drkappa} \\
k=4\hspace{0.8cm} b_{0,0}-\frac{\kappa_3^2}{ M_{P}^6}t+{O}(t^2) &=& \Big\langle \frac{d^J_{0,0}(1+2t/m^2)}{(m^2)^4} \Big\rangle^{g_{++}^2} + \Big\langle \frac{m^2\,d^J_{4,4}(1+2t/m^2)}{(m^2+t)^5} \Big\rangle^{g^2_{+-}} \,.
\label{drcase1}
\eea
From the $k=2$ sum rule \eq{drmp}, 
we see that both terms on the RHS, $s$ and $u$ channel contributions, appear
with positive sign.
 However, we cannot take the limit $t\to0$ since in this limit the dispersion relations are not convergent. Indeed, the LHS of \eq{drmp} contains  the pole $1/t$ induced by the exchange of a graviton, which cannot be matched to the RHS of the dispersion relations that only have powers of $t$.
We could consider \eq{drmp} at finite $t$, but  the  Wigner $d$-functions at fixed $t$ are not positive for all $m$ and $J$. 
The  strategy  to overcome this problem  is  to integrate over $t$ the sum rules 
weighted by some  particular functions that can guarantee the positivity of the high energy averages.
This technique is referred to as \emph{smearing} the sum rules
\cite{Caron-Huot:2021rmr, Caron-Huot:2022ugt, Caron-Huot:2022jli,Beadle:2024hqg}.

Taking $k_{\rm min}=2$, as justified in   \cite{Caron-Huot:2022ugt},
a bound on $\kappa_g$ will be derived in the following way. 
First, we  convolute the $k=2$ and $k=3$ dispersion relations \eq{drmp} and \eq{drkappa} respectively with   the smearing functions $A(t)$ and $B(t)$:
\bea
\int^0_{-|t|_\text{max}} dt \frac{-A(t)}{M_P^2\ t} &=& \sum_i^{s\text{-ch}}A_s(m_i,J_i) g_{++,i}^2 + \sum_j^{u\text{-ch}}A_u(m_j,J_j) g^2_{+-,j}\, , \label{drmp2}  \\
-\int^0_{-|t|_\text{max}} dt B(t)
\bigg(\frac{\kappa_g^2}{M_P^4 F_\pi^2} +\frac{\kappa_3^2}{ M_{P}^6}t^2\bigg)
&=& \sum_i^{s\text{-ch}}B_s(m_i,J_i) g_{++,i}^2 + \sum_j^{u\text{-ch}}B_u(m_j,J_j) g^2_{+-,j}\, , \label{drkappa2}
\eea
where we  used the tree-level approximation in \eq{eq:imM} to rewrite the RHS as an infinite sum over states and we defined the functions 
\bea
A_s(m,J)= \int_{-|t|_\text{max}}^0 dt A(t) \frac{d^{J}_{0,0}(1+\frac{2t}{m^2})}{m^4} &,&  B_s(m,J)= \int_{-|t|_\text{max}}^0 dt B(t) \frac{d^{J}_{0,0}(1+\frac{2t}{m^2})}{m^6}\, , \hspace{1.8cm} \nonumber \\ 
 A_u(m,J)= \int_{-|t|_\text{max}}^0 dt A(t) \frac{m^2d^{J}_{4,4}(1+\frac{2t}{m^2})}{(m^2+t)^3} &,& B_u(m,J)= -\int_{-|t|_\text{max}}^0 dt B(t) \frac{m^2d^{J}_{4,4}(1+\frac{2t}{m^2})}{(m^2+t)^4}\, . \label{aubu}
\eea
We integrate over $t$ negative so that we are guaranteed not to encounter any $t$-channel pole.
 We will specify later  the value of $|t|_\text{max}$ from which we integrate.

Next, we look for $A(t),B(t)$ such that each term on the RHS of \eq{drmp2} is greater than or equal to the corresponding term on the RHS of \eq{drkappa2}:
\bea
A_s(m_i,J_i) &\geq&B_s(m_i,J_i) \hspace{1.4cm} \forall m_i^2 \geq M_s^2 \text{ \ \  and \ \  }\forall J_i\geq 0  \, ,\label{relationss} \\
A_u(m_i,J_i) &\geq&B_u(m_i,J_i) \hspace{1.4cm} \forall m_i^2 \geq M_u^2 \text{ \ \  and \ \   }\forall J_i \geq 4 \, , \label{relationsu}
\eea 
where $M_s$ and $M_u$ is respectively the mass of the lightest resonance encounter in the $s$
and $u$-channel (that due to helicity conservation must have  $J\geq 4$). 
Using the conditions \eq{relationss} and \eq{relationsu}  on the RHS of \eq{drmp2} and \eq{drkappa2}, we can  obtain  a bound on $\kappa_g^2$:
\be
-\int_0^{|t|_\text{max}} dt B(-t)
\bigg(\frac{\kappa_g^2}{M_P^4 F_\pi^2} +\frac{\kappa_3^2}{ M_{P}^6}t^2\bigg)
\leq
\int_0^{|t|_\text{max}} dt {A(-t)}/{t}
\, ,
\label{lerelations}
\ee 
where we have changed the integration variable  $t\to -t$ and flipped the integration limits.
The bound \eq{lerelations} gives  a non-trivial upper bound on $\kappa_g$ only if  the LHS is positive, so we must also demand that $B(t)$  satisfies
\be 
\int_0^{|t|_\text{max}} dt B(-t)\leq 0 \qquad \text{and} \qquad \int_0^{|t|_\text{max}} dt B(-t) t^2 \leq 0\, .
\label{req2}
\ee
To maximise the bound \eq{lerelations}, we must take $|t|_\text{max}$ to be as large as possible. Nevertheless, we cannot take  $|t|_\text{max}\geq M_u^2$, otherwise $A_u(m,J)$ and $B_u(m,J)$ in \eq{aubu} will have  singularities at  $t\to -m^2$ and the inequalities in \eq{relationsu} would not be  satisfied. 
We will therefore take
\be
|t|_\text{max}=M_u^2\,,
\ee
and choose  smearing functions such that the only singularities at  $t=-M_u^2$ 
in $A_u(m,J)$ and $B_u(m,J)$ are properly cancelled.

\begin{figure}[t]
\centering
\hskip-.5cm  \includegraphics[width=0.48\textwidth]{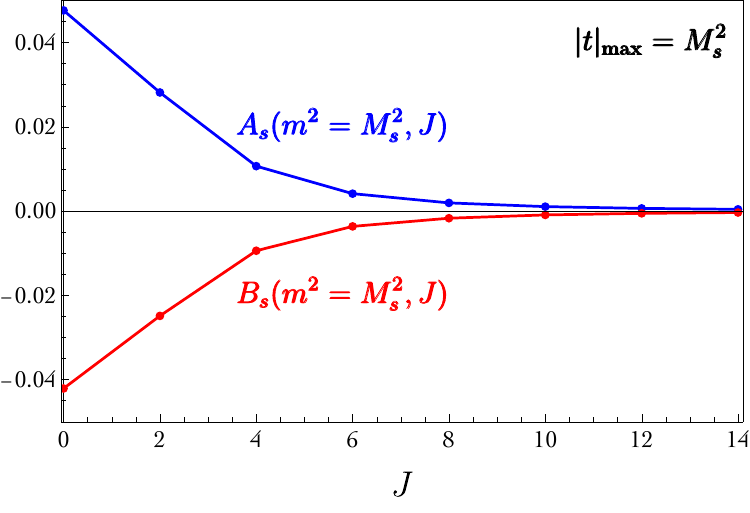}
\includegraphics[width=0.47\textwidth]{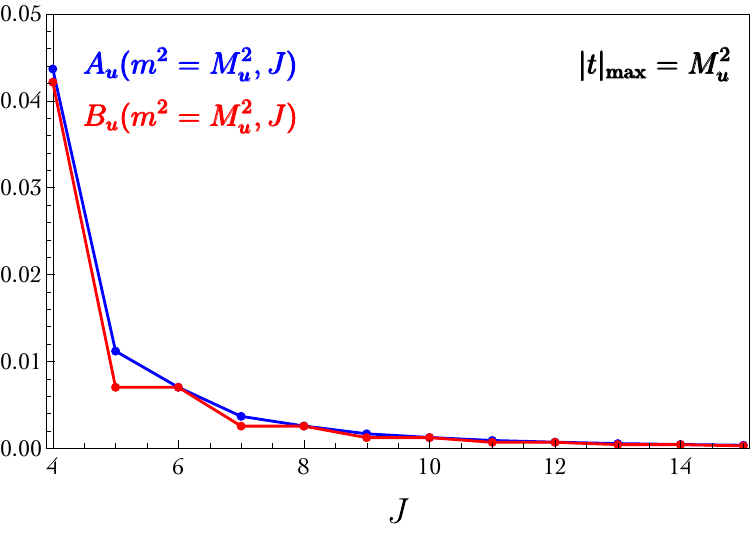}
\caption{\it 
Values of $A_{s,u}$ and $B_{s,u}$, defined in \eq{aubu}, for the smearing functions \eq{functions}.
We have taken $m_i$ equal to its lowest value, $m_i=M_s\ (M_u)$ in  $A_s,B_s$ ($A_u,B_u$). 
}
\label{fig:asbsaubu}
\end{figure}

An example of smearing functions that  satisfy all the  requirements mentioned above
are 
\be
A(t) = (1-\sqrt{-t/|t|_\text{max}})^5\quad \text{and}\quad B(t)=-0.885 |t|_\text{max}(1-\sqrt{-t/|t|_\text{max}})^5 \,.
\label{functions}
\ee
We have checked explicitly that  the  functions above satisfy the inequalities \eq{relationss} and  \eq{relationsu} for
general $m_i$ and $J_i$. 
For the lowest value  $m_i=M_{s,u}$,\footnote{It is sufficient to fix the masses to their lowest possible values, $m_i=M_{s,u}$, and check that the inequalities are satisfied for all possible values of $J_i$. 
If so, then they will also be satisfied for larger masses.} 
we show $A_{s,u}$ and  $B_{s,u}$  in Fig.~\ref{fig:asbsaubu} as a function of $J_i$.
As we increase $J_i$ the values of $A_{s,u}$ and  $B_{s,u}$  are difficult to get numerically, but we  show in
 Appendix~\ref{applargeJ} that  
 the conditions \eq{relationss} and  \eq{relationsu} can be satisfied for $J\to\infty$.
 In this limit the smeared functions decay like $1/J^3$, 
 property which we derive analytically in Appendix~\ref{applargeJ}, but also observe in Fig.~\ref{fig:asbsaubu}.
The relations \eq{relationss} and  \eq{relationsu} are also satisfied when the masses of the states in the $s$ channel are smaller than  $|t|_\text{max}$, i.e, for  $M_s<M_u$.

We notice that  the functions  \eq{functions} also satisfy
\be
A_{s,u}(m_i,J_i)\geq0\,,
\label{positivity}
\ee
 indicating that  each  term of  \eq{drmp2}  is positive.  This shows that the sum rule \eq{drmp2}
is exactly what we were looking for, an $O(1/m^4)$ positive sum rule.

Plugging the smearing functions \eq{functions} into \eq{lerelations}, we find the bound 
\be
\frac{M_u^4 \kappa^2_g}{M_P^2 F^2_\pi}+0.015\frac{\kappa_3^2 M_u^{8}}{ M_{P}^4} 
\lesssim  48\ln\bigg(\frac{M_u}{M_\text{IR}}\bigg)\,,
\label{firstbound0}
\ee
where $M_\text{IR}$ is an IR-cutoff that we are forced to introduce to regularize the logarithmically divergent integral  $\int dt {A(-t)}/{t}$.
This is a well-known fact of 4D gravity that also appears
in constraints from  time delay in the eikonal limit \cite{Camanho:2014apa}.  
We must keep $M_\text{IR}$  as an independent parameter
but we can assume to take values close to $M_u$ such that $\ln({M_u}/{M_\text{IR}})\sim O(1)$.
Although the bound  \eq{firstbound0} could be improved by a better selection of  smearing functions \cite{Caron-Huot:2022ugt}, we have checked that the result
does not  change substantially.
Since both terms on the LHS of \eq{firstbound0} are positive, we can derive the bound
\be
\left|\frac{M_u^2 \kappa_g}{M_P F_\pi}\right| 
\lesssim  7\ln^{\frac{1}{2}}\bigg(\frac{M_u}{M_\text{IR}}\bigg)\, .
\label{firstbound}
\ee
We recall that $M_u$ in \eq{firstbound} corresponds to the mass of the lightest state of spin $4$ or higher.
Therefore \eq{firstbound} can be interpreted 
as a bound 
due to the presence of the anomaly, $\kappa_g\not=0$,
on the mass
of   states with   $J\geq 4$  and nonzero coupling to two gravitons, $g_{+-,i}\not=0$. 
These   states are of  type $n=1,2$  as defined in Table~\ref{tab:states}.
Similarly, a bound on $M_u$ can also be derived from \eq{firstbound} even when $\kappa_g=0$ as long as  
$\kappa_3\not=0$ (this is similar to the bound in \cite{Caron-Huot:2022ugt}). Nevertheless, $\kappa_3\not=0$  is not necessarily guaranteed by any intrinsic UV condition of the model such as the anomaly.

There are other ways to bound
$\kappa_g$ besides the one given above. An alternative  possibility
is shown in  Appendix \ref{2ndbound} where we obtain  a bound on $\kappa_g$ 
from  the Wilson coefficient $b_{0,0}$:
\be
\frac{M_u \kappa_g^2}{M_P^3 F_\pi^2} + 0.023 \frac{M_u^5 \kappa_3^2}{M_P^5} \lesssim 21 \sqrt{b_{0,0}} \ln^{\frac{1}{2}} \bigg(\frac{M_{u}}{M_\text{IR}}\bigg)\,.
\label{boundwilson}
\ee

\subsection{Eta-graviton scattering}
\label{etagrav}

Similar bounds as the ones found above can be obtained from amplitudes involving external gravitons and 
pseudo-scalars $\eta$    shown in Fig.~\ref{fig:diagrams2}. 
As we will see below, the bound derived  in these cases will  be  parametrically similar to \eq{firstbound}.

Below we will derive the simplest bound from  the elastic process $\eta h^+ \rightarrow \eta h^+$ (B in Fig.~\ref{fig:diagrams2}), while  in Appendix \ref{elimit} we will also show how to construct a bound from the process $\eta h^+ \rightarrow h^+ h^+$ (C in Fig.~\ref{fig:diagrams2}) using a trigonometric inequality. The same logic could be also  used to obtain  bounds from  $\eta h^+ \rightarrow \eta \eta $ (D in Fig.~\ref{fig:diagrams2}), which however we will  not discuss because  they are  similar to the ones derived here. We will  show in Appendix \ref{eiksub} 
how in  these cases  the eikonal limit can be recovered leading to a similar  bound as the one derived  in \cite{Caron-Huot:2022ugt} for $\kappa_3$.

The amplitude for the process $\eta h^+ \rightarrow \eta h^+$
at low energy is given by  
\be
{\cal M}_{\eta+\eta-}\rightarrow s^2 u^2\bigg(-\frac{1}{M_P^2 s t u}-\frac{\kappa_g^2 t}{M_P^4F_\pi^2 su} +...\bigg)\, .
\label{lowenhnh}
\ee
At fixed $t$, we use the dispersion relation $\oint_{C_\infty} M_{\eta+\eta-}/s^2u=0$ instead of $\oint_{C_\infty} M_{\eta+\eta-}/s^3=0$ (both integrants have the same high-energy behaviour) 
to avoid the appearance of Wilson coefficients. We find
\be
\frac{-1}{M_P^2t} -\frac{\kappa_g^2 t}{M_P^4F_\pi^2} =
\Big\langle\frac{d_{-2,-2}^J(1+2t/m^2)}{m^2(m^2+t)}
\Big\rangle^{g_{\eta +}^2}
+
\Big\langle\frac{d_{2,2}^J(1+2t/m^2)}{(m^2+t)^2}
\Big\rangle^{g_{\eta-}^2}\,.
\label{dranother}
\ee
Since ${\cal M}_{\eta+\eta-}$ is $s\leftrightarrow u$ symmetric, we have $M_u=M_s$. 
Integrating \eq{dranother} in $t$ with a smearing  function $C(t)$,  we get  
\be
\int^0_{-|t|_\text{max}}C(t) \bigg(\frac{-1}{M_P^2t} -\frac{\kappa_g^2 t}{M_P^4F_\pi^2} \bigg)dt= \bigg(\sum_i  C_s(m_i,J_i) g_{\eta+,i}^2 + \sum_j C_u(m_j,J_j) g^2_{\eta-,j}\bigg)\, ,
\label{smdispelastic}
\ee
where we defined the functions 
\bea
C_s(m,J)&=& \int_{-|t|_\text{max}}^0 C(t) \frac{d_{-2,-2}^J(1+2t/m^2)}{m^2(m^2+t)} \ dt \nonumber \, ,\\
C_u(m,J)&=& \int_{-|t|_\text{max}}^0 C(t) \frac{d_{2,2}^J(1+2t/m^2)}{(m^2+t)^2}\ dt \, .
\label{Cdef}
\eea
Demanding that the  smearing function $C(t)$ satisfies
\be
C_{s,u}(m,J) \geq0   \hspace{1.5cm} \forall m^2\geq M_{s}^2 = M_{u}^2\ , \ \forall J\geq 2\, , 
\ee
we can use \eq{smdispelastic} to obtain 
\be
\int_0^{|t|_\text{max}}C(-t) \bigg(\frac{1}{M_P^2t} +\frac{\kappa_g^2 t}{M_P^4F_\pi^2} \bigg)dt\geq 0\, .
\label{uvconstrnhnh}
\ee
To find a non-trivial bound  on $\kappa_g^2$  from the equation above
we need  the first and second term to have  opposite  signs.
A smearing function that satisfies these conditions and also guarantees the positivity requirement \eq{uvconstrnhnh} is 
\be
C(t)= J_0(\sqrt{- \beta^2 t/|t|_\text{max}}) \left(1-\sqrt{-t/|t|_\text{max}}\right)^3\  \ \ (\beta\geq 5)\,,
\label{cc}
\ee
where $J_0$ is the zero-order Bessel function. To 
 maximize the bound we take\footnote{This is the largest  value of  $|t|_\text{max}$ such that   the integrals in \eq{Cdef} do not hit the singularities at  $t\to -m^2$.}
   \be
 |t|_\text{max}=M_s^2=M_u^2\equiv M_2^2\,,
 \label{m2def}
  \ee
where $M_{2}$ corresponds to  the mass of the lightest    $J\geq2$ state with $g_{\eta\pm,i}\not=0$.
These   states are of  type $n=1,2,3$  as defined in Table~\ref{tab:states}.
Using \eq{m2def}   in \eq{cc} and setting  $\beta \sim 7.2$ (this value maximizes the bound),  
  \eq{uvconstrnhnh} yields
\be
\frac{M_{2}^2 \kappa_g}{M_P F_\pi} \lesssim 30 \ln^{\frac{1}{2}} \bigg(\frac{M_{2}}{M_\text{IR}}\bigg)\, .
\label{thirdbound}
\ee

Bounds can also be derived from  inelastic processes. 
In Appendix \ref{elimit} we use  the process 
 $\eta h^+ \rightarrow h^+ h^+$ to  obtain the bound
\be
\left|\frac{M_2^2 \kappa_g}{M_P F_\pi}-0.017\frac{M_2^6\kappa_3\kappa_g}{M_P^3F_\pi} \right|\lesssim 106 \ln\bigg(\frac{M_2}{M_\text{IR}}\bigg)\,.
\label{secondbound}
\ee
The second term on the LHS, which can have either sign, seems to allow  $\kappa_g$ to be unbounded.  
Nevertheless, the numerical coefficient in front of this second  term could have been made much smaller  by a better selection of the smearing function.
In fact, as discussed in Appendix~\ref{eiksub}, it goes to zero in the  eikonal limit.

\section{Phenomenological Implications}
\label{pheno}

The bounds \eq{firstbound} and \eq{thirdbound} 
can  be  interpreted as  an upper bound for $M_u$ and $M_2$ respectively in theories with $\kappa_g\not=0$.
Since both bounds are almost  identical, we will only consider here the implications
for the upper bound on $M_u$.

This upper bound represents  the  mass scale at or below which $J\geq 4$ states must appear in 
${\cal L}_{\rm EFT}(\eta,h)$,  the EFT  of  $\eta$  and gravitons.
This mass scale, which from now on we will refer to as  $\Lambda_{\text{caus}}$,  
represents  then  a cutoff scale for ${\cal L}_{\rm EFT}(\eta,h)$.
Taking  $M_{\rm IR}= O(M_{u})$     in \eq{firstbound},   we find that  this  is parametrically given by
\be 
 \Lambda_{\text{caus}}\sim \sqrt{\frac{M_PF_\pi}{\kappa_g}}\,.
\label{cutoff}
\ee
For  $\kappa_g\sim 1$, $\Lambda_{\text{caus}}$ lies around the geometric mean of $M_P$ and $F_\pi$.
This bound was also found in \cite{Serra:2022pzl} from time delay analysis, and was used to
put interesting constraints on theories beyond   General Relativity.

The  cutoff scale $\Lambda_{\text{caus}}$ is smaller than $\Lambda_{\text{pert}}$, 
the naive perturbative cutoff scale
of ${\cal L}_{\rm EFT}(\eta,h)$  defined as the scale at which loops become of order tree-level.
For ordinary gravity,   $\Lambda_{\text{pert}}$ is around $M_P$, but for ${\cal L}_{\rm EFT}(\eta,h)$  with the coupling \eq{etahh}, this is given by\footnote{This can be estimated by the energy scale at which the $\kappa_g^2$ contribution to $M_{++++}$  (first term of \eq{eq:pppp}) becomes $O(1)$.}
\be 
\Lambda_{\text{pert}}\sim  \sqrt[3]{\frac{M_P^2F_\pi}{\kappa_g}} \,.
\label{pertbound}
\ee
For $F_\pi\lesssim \kappa_gM_P $, we have indeed 
that $\Lambda_{\text{caus}}\lesssim \Lambda_{\text{pert}}$.

\subsection{Models of axions}

Axions  arise as  pseudo-scalar Goldstone bosons of  non-linearly realized   global $U(1)$ symmetries.
In the case of the QCD axion,  the  $U(1)$ has 
a  $U(1)-SU(3)_c-SU(3)_c$ anomaly  that induces  the interaction term
${\cal L}_{\rm int}=aG\tilde G/f_a$, being $G$ the $SU(3)_c$ field-strength and 
 $f_a$  the axion decay constant.
It is   natural to expect that these models also have a $U(1)$-gravitational anomaly (as is the case in most proposed models),  leading to  the presence of \eq{etahh}   with $\eta$ identified as the axion $a$ and $F_\pi/ \kappa_g\sim f_a$.
In this case, \eq{cutoff} tells us that the  EFT of  axions  cannot be extrapolated beyond 
\be
\Lambda_{\text{caus}}\sim \sqrt{M_Pf_a}\,,
\label{axionb}
\ee
where new states must appear.
This cutoff,   expected   to be present  in any  axion-like EFT, 
is smaller than  any other cutoff scale based on perturbativity  such as \eq{pertbound}.

In explicit weakly-coupled UV realizations of  axion models, new physics indeed appears below 
\eq{axionb}. For instance,  in axion  models in which  $\kappa_g$ is one-loop generated by  massive extra quarks, 
the mass of these states is  always below \eq{axionb}.\footnote{To understand how the positivity bound is relaxed above the quark mass, see \cite{Bellazzini:2021shn} for a similar case.}

\subsection{Models  of massive higher-spin resonances}

Let us now consider  the class of models consisting  of 
an infinite set of states $R_i$ with arbitrary spin $J$
and a characteristic coupling $g_*$. 
We will be considering the weak-coupling limit $g_*\to 0$.
More realistic scenarios  containing different sets of states, each with its own distinct $g_*$, will be introduced later.

We couple this class of models to gravity.
We can schematically write the Lagrangian of the states $R_i$  interacting with  gravity as
\be  
{\cal L}=M_P^2 {\cal R}+ \frac{1}{g^2_*} \left[(\partial R_i)^2 + m^2_i R^2_i+m_i R_i^3+ h_{\mu\nu}T^{\mu\nu}(R_i)+\cdots\right]\,,
\label{lagrangianR}
\ee
where ${\cal R}$ is the Ricci scalar that contains the kinetic term of the graviton and 
$T^{\mu\nu}(R_i)$ is the energy-momentum tensor  that can be expanded as  a polynomial in $R_i$.
Assuming that the interactions of $\eta$ have the same parametric dependence as those of $R_i$
given in \eq{lagrangianR}, we can   estimate
\be
F^2_\pi\sim M^2/g^2_*\ , \ \ \ \kappa_g\sim 1/g^2_*\,,
\ee
where $M$ is the mass of the lightest resonance,  in principle,  not necessarily a $J\geq 4$ state that could be much heavier.

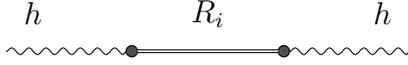
\begin{figure}[t]
\centering
  \begin{tikzpicture}[baseline=.1ex]
        \begin{feynman}[every dot={/tikz/fill=black!70}]
            \vertex[dot] (m) at (1, 0){};
            \vertex[dot] (e) at (-1,0) {};
            \vertex (a) at (-2.8,0) {$$} ;
            \vertex (d) at ( 2.8, 0) {$$};
        \vertex (f) at (-2.3,0.5) {$h$} ;
            \vertex (g) at ( 2.3, 0.5) {$h$};
           \vertex (h) at ( 0, 0.5) {$R_i$};
            \diagram* {
                (a) -- [photon] (e) --[double] (m) -- [photon] (d),
 
            };
            \end{feynman}
            \end{tikzpicture}
            \caption{\it 
Contribution to the graviton propagator due to the exchange of resonances.}
\label{fig:gravvertex}
\end{figure}

One could guess that in the formal limit $g_*\to 0$ and  $M_P/M\to \infty$,   the model is valid at all energies.
Nevertheless, this is not  true. 
From the interchange of resonances as in Fig.~\ref{fig:gravvertex},
the graviton 2-point function $\Pi(q)=q^2$  (stripping off the polarization tensor)
 receives corrections that,  on dimensional grounds, can be estimated to be  
 \be
\Delta \Pi(q)\sim \frac{q^4}{g^2_*M_P^2} \,.
\ee
These corrections can overcome the tree-level  contribution for momentum $q$ above  the scale
\be
 g_* M_P\equiv \Lambda_{\rm QG}\,,
\label{qgscale}
\ee
that therefore corresponds to a UV cutoff scale in the theory.
Although we determined the cutoff scale $\Lambda_{\rm QG}$ from naive perturbativity,
it has been recently argued that this cutoff scale  is rigorous and 
 related  with  UV completions of gravity \cite{Caron-Huot:2024lbf}. 
Therefore, we expect at  $\Lambda_{\rm QG}$ states associated to quantum gravity,
such as string modes,  
which are unrelated to the states in our model of resonances.

We will work in the limit
\be
 M_P\to \infty\ , \ \ \
 g_*\to  0\ , \ \ \
 \Lambda_{\rm QG}={\rm Fixed}\,.
 \ee
In this limit, the bound  based on causality \eq{cutoff} goes as
\be
\Lambda_{\text{caus}}\sim \sqrt{g_* M_P M}\,,
\label{largencausa}
\ee
that using \eq{qgscale} can be written as 
\be
 \Lambda_{\text{caus}}\sim \sqrt{\Lambda_{\text{QG}} M}\,.
 \label{causabound}
\ee
Since   $\Lambda_{\text{QG}}\gg M$,  
the scale below which  spin$\geq4$  states must appear is always smaller than the   scale of quantum gravity.
This  is therefore an important bound  for these type of  models when $\kappa_g\not=0$.

\subsubsection{Strongly-coupled large-$N_c$ gauge theories}
\label{largeNgauge}

Models with similar properties as the ones described above are 
strongly-coupled gauge theories with $N_F$ fermions ($f_{L,R}$) in the $N_c\to \infty$  limit. In this section
we will  consider that the 'tHooft coupling, $g^2_{\rm YM}N_c$, is not large, and then
physical quantities are characterized only by  $N_c$ and $M$, the mass gap of the theory \cite{tHooft:1973alw, Witten:1979kh}. 

The pseudo-scalar $\eta$ can appear in these theories as  the Goldstone boson associated to the 
chiral breaking $U(1)_L \times U(1)_R\to U(1)_V$.
In principle, since the axial $U(1)_A$ current $J^\mu_A = \sum_f\bar{f} \gamma^\mu \gamma^5 f$ is not conserved at the quantum level due to the axial anomaly,   $\eta$ gets a mass
of order $m^2_\eta\sim N_F M^4/F_\pi^2$ \cite{Witten:1979vv, Veneziano:1979ec}. 
In the large-$N_c$ limit, however,
 $m^2_\eta\to 0$   since $F^2_\pi\sim N_c M^2\to \infty$.
Unlike the models described above, 
 large-$N_c$   gauge theories  contain two distinct sets of resonances, mesons and glueballs, whose   3-point couplings
 scale differently at large $N_c$:
\be
g_{\rm meson}\sim \frac{1}{\sqrt{N_c}}\ , \ \ \ \
g_{\rm glueball}\sim \frac{1}{N_c}\,.
\label{largeNcouplings}
\ee

By switching on the gravitational interactions, the  coupling  \eq{etahh} is generated due to the $U(1)$-gravitational anomaly,  with  $\kappa_g \sim N_c\sqrt{N_F}$.
For the particular case  of $SU(N_c)$ gauge theories, we have 
\be
 \kappa_g =\frac{N_c\sqrt{N_F}}{192\pi^2}\,.
\label{anomaly}
 \ee
Corrections to the graviton propagator from 
Fig.~\ref{fig:gravvertex} are dominated by glueballs, leading at $q\gg M$ to
 \be
\Delta \Pi(q)\sim \frac{N_c^2 q^4}{M_P^2}\ln q \,.
\ee
This correction can also be understood as arising at high energy from a loop of gluons. 
Therefore we can define 
\be
\Lambda_{\rm QG}=\frac{M_P}{N_c}\,,
\label{qgscale2}
\ee
as the scale where the corrections dominate over tree-level.
Using these relations in \eq{cutoff}, we get
\be
 \Lambda_{\text{caus}}\sim \sqrt{\Lambda_{\text{QG}} M\sqrt{\frac{N_c}{N_F}}}\,.
\label{boundncnf}
\ee
In the  strict  $N_c\to \infty$ limit  at $N_F$ fixed,  we see that $\Lambda_{\text{caus}}$ overcomes  
 $\Lambda_{\text{QG}}$ and no limit can be deduced for the $J\geq4$ resonances of the gauge sector.
 This is due to the presence of the glueballs in Fig.~\ref{fig:gravvertex}  that gave corrections to the graviton propagator
 of order $N^2_c$,  larger than meson contributions that are of  $O(N_c)$.

Notice however that for theories in the Veneziano limit,  $N_c\to \infty$  with $N_c/N_F$ fixed,  the bound \eq{boundncnf}  remains finite, predicting that $J\geq 4$ resonances must be present below
$\Lambda_{\text{QG}}$. 

Our analysis can also be used to put bounds on 3-point couplings. For example, \eq{firstbound0} leads  to a bound
on $\kappa_3$ as a function of the  mass of the lightest $J\geq 4$ state, $\Lambda_{\rm caus}$,
 that parametrically is given by
 \be
|\kappa_3|\lesssim  \frac{M^2_P}{\Lambda^4_{\rm caus}}=N_c^2\left(\frac{\Lambda_{\rm QG}}{\Lambda_{\rm caus}}\right)^2
\frac{1}{\Lambda^2_{\rm caus}}\,.
\label{boundk3}
\ee
Since $\kappa_3$ can be generated by the exchange of glueballs (loops of gluons), it scales as $N^2_c$, and therefore
 \eq{boundk3} provides a nontrivial upper bound in the limit $N_c\to\infty$. 
 If we require $\Lambda_{\rm caus}\to \infty$ (with $\Lambda_{\rm caus}/\Lambda_{\rm QG}\ll 1$ fixed), we get that $\kappa_3\to 0$.


In theories with a light   spin-0 glueball (with a mass $\leq M$),  we can also obtain a bound on its coupling to gravitons. 
This bound is easy achieved by repeating the above  derivation that led to \eq{cutoff}  but replacing the pseudo-scalar $\eta$ with   the scalar glueball. We  obtain  again \eq{cutoff} but with the replacements
\be
 F_\pi\to F_{\rm glueball}\sim N_c M\ , \ \ \ \      \kappa_g\to  \kappa_{\rm glueball}\sim N_c^2\,,
\ee
where $\kappa_{\rm glueball}$ is the coupling of the glueball to two gravitons,  that leads to   the parametric  bound
\be
| \kappa_{\rm glueball}|\lesssim \frac{M_P F_{\rm glueball}}{\Lambda^2_{\rm caus}}\sim 
N^2_c\,\frac{\Lambda_{\rm QG} M}{\Lambda^2_{\rm caus}} \,. 
\label{newbound}
\ee
Therefore in  models in which $J\geq 4$ states are very heavy, $\Lambda_{\rm caus}\gg M$, one  must have
 $\kappa_{\rm glueball}/N_c^2\to 0$. This is  especially relevant for models with a holographic dual as we will see in the next section. 

Interestingly, in models where the light glueball is a dilaton, the Goldstone associated to the  spontaneous breaking of scale invariance  in a conformal  field theory (CFT) (see next section for a model like this),
we have \cite{Karateev:2023mrb}
\be
\kappa_{\rm glueball}=\kappa_{\rm dilaton}={\sqrt{2}} (a_{\rm UV}-c_{\rm UV})\,,
\label{dilatoncase}
\ee
where $a_{\rm UV}$ and $c_{\rm UV}$ are  the coefficients of the $c$- and $a$-trace anomalies respectively (we have $a_{\rm IR}=c_{\rm IR}=0$ as we are considering gapped theories at the IR).
Therefore our bound \eq{newbound} provides a constraint on $(a_{\rm UV}-c_{\rm UV})$.
For $\Lambda_{\rm caus}\gg M$, we get $( a_{\rm UV}-c_{\rm UV})/N^2_c\ll 1$.

We could also consider the dilaton to be external to the strongly-coupled CFT, as in  \cite{Karateev:2023mrb}.
In this case  $F_{\rm dilaton}$ is independent of the strong sector and does not scale with $N_c$.
We  must however specify how to take the weak-coupling limit $F_{\rm dilaton}\to \infty$ 
with respect to $N_c\to \infty$.
As for the graviton case, we must   guarantee that the tree-level  propagator of the external dilaton is not affected by the exchange of  glueballs  (Fig.~\ref{fig:gravvertex} with the graviton replaced by the external dilaton).
This defines a new cutoff scale (similarly as \eq{qgscale2})
\be
\frac{F_{\rm dilaton}}{N_c}\equiv \Lambda_{\rm QD}\,,
\ee
that we take fixed as $F_{\rm dilaton}\to \infty$ and $N_c\to\infty$.
We obtain  the  parametric bound
\be
\frac{|a_{\rm UV}-c_{\rm UV}|}{N^2_c}\lesssim \frac{\Lambda_{\rm QG} \Lambda_{\rm QD}}{\Lambda^2_{\rm caus}} \,. 
\label{newbound2}
\ee
Since the last ratio must be larger than one, this does not provide  a very stringent constraint.

\subsubsection{Holographic AdS$_5$ models}

Another  class of  models with similar properties  are  
holographic  theories based on the AdS/CFT correspondence.
These are weakly-coupled models in more than 4D  conjectured to be the duals of strongly-coupled models 
in the large-$N_c$ limit.
The most famous example is the 10D AdS$_5\times S^5$ model dual to ${\cal N}=4$ supersymmetric Yang-Mills theories \cite{Aharony:1999ti}.  
In the large 'tHooft coupling limit, $\lambda\equiv g_{\rm YM}^2N_c\to \infty$, the 
$J>2$ states  (associated to string states)
of the dual 10D theory   become infinitely heavy, and the light spectrum consists of 
states of $J=0,1,2$.
It is then interesting to  understand whether  the cutoff 
$\Lambda_{\rm caus}$ derived above imposes  constraints on this class of models.
 
We will consider here a simple  5D model  that at low-energies consists of a 4D massless $\eta$ and a graviton.
The 5D Lagrangian is given by (up to marginal operators)
\be
{\cal L}_5=\int d^5x \sqrt{-g}  \left[M_{P5}^3 {\cal R}+M_5 F_{MN}^2+\kappa_5\, \epsilon_{MNPQR} A^M {\cal R}^{NPA}_B {\cal R}^{QRB}_A\right]\,,
\label{lagrangian5}
\ee
where $F_{MN}$, ${\cal R}$ and ${\cal R}^{MNPQ}$
are respectively the gauge field-strength, Ricci scalar and Riemann tensor;
$M_{P5}$  and $1/M_{5}$ are respectively the 5-dimensional Planck scale and 
 gauge-coupling squared, and $\kappa_5$ is the dimensionless coefficient of  the Chern-Simons 
term related with the  chiral-gravitational anomaly.

We assume the metric is AdS$_5$, $ds^2=(L/z)^2 (\eta_{\mu\nu}dx^\mu dx^\nu-dz^2)$, being $L$  the AdS curvature radius and 
$z$ the 5th dimension.
We also consider that the  extra dimension  is compactified  by placing   two branes, one at $z =1/\Lambda_{\rm UV}$  (UV-boundary) and another at $z =1/\Lambda_{\rm IR}$  (IR-boundary)
with $\Lambda_{\rm UV}\gg \Lambda_{\rm IR}$.
The model is then defined on the line segment $1/\Lambda_{\rm UV} \leq z \leq 1/\Lambda_{\rm IR}$.
 This set-up is referred as a Randall-Sundrum model \cite{Randall:1999ee}.
 
As boundary conditions  we choose $A_\mu=0$ at  both boundaries such that at low-energy only the fifth-component 
of the gauge boson  $A_5$ (a pseudo-scalar) remains in the spectrum. 
At low-energy $E\ll \Lambda_{\rm IR}$ 
then the model has two massless states,  $A_5$ to be identified with $\eta$ and
the zero-mode of the graviton $h_{\mu\nu}$. 
 It is easy to check that the Chern-Simons term  leads to \eq{etahh}  with $\kappa_g\propto \kappa_5$.
If we tune the brane tensions as in \cite{Randall:1999ee}, we  also get a massless scalar from the gravitational sector, the radion, that corresponds to a dilaton.
This however gets a mass  when the branes are not tuned but  dynamically stabilized \cite{Goldberger:1999un}.

The  model contains an infinite set of  massive resonances corresponding 
to the Kaluza-Klein (KK) modes of the graviton and gauge boson with masses $\sim r\Lambda_{\rm IR}$ where $r\in\mathbb{N}^{+}$.
   Notice that the model only contains $J=1,2$ KK states. 
Their 3-point coupling squared for KK graviton and gauge bosons are parametrically given by
\be 
g^2_{\rm KK, grav}\sim 1/(M_{P5}L)^3\ ,\ \  \ \ 
g^2_{\rm KK, gauge}\sim 1/(M_{5}L)\,.
\label{couplingsKK}
\ee
Therefore  by working  in the limit 
\be
(M_{P5}L)^3\to \infty\ ,\ \ \    M_{5}L\to \infty \,,
\label{hololimits}
\ee
the couplings of the model go to zero becoming a free theory.
It is easy to calculate $F_\pi$, $M_P$ and $\kappa_g$ for  the EFT of the 4D massless modes. One gets
\be
F_\pi^2=M_5L\Lambda_{\rm IR}^2\ , \ \ \ \
M_P^2=(M_{P5} L)^3\Lambda_{\rm UV}^2\ , \ \ \ \ \kappa_g=\kappa_5 \,.
\label{holorelations}
\ee

 In the limit \eq{hololimits},
the only UV cutoff of the model  is given   by $\Lambda_{\rm UV}$. 
Indeed, this scale can be identified with  $\Lambda_{\rm QG}$ by looking again to the corrections of 
the graviton propagator \cite{Arkani-Hamed:2000ijo}:
\be
\Delta \Pi(q)= \frac{(M_{P5}L)^3}{M_P^2}  q^4 \ln q= \frac{1}{\Lambda_{\rm UV}^2}  q^4 \ln q\,,
\ee
that becomes larger than the tree-level contribution  above  $\Lambda_{\rm QG}\sim \Lambda_{\rm UV}$.

Plugging \eq{holorelations} into  \eq{cutoff},  we  obtain that this theory becomes acausal at energies above
\be
\Lambda_{\rm caus}\sim r_5 \sqrt{\Lambda_{\rm UV} \Lambda_{\rm IR}}\ , \ \ \ \ \ 
r_5\equiv \left(\frac{M_{P5}^3 M_5L^4}{\kappa_5^2} \right)^{1/4}\,.
\ee
If we demand this scale to be bigger than $\Lambda_{\rm UV}$, in order to avoid   problems with causality 
below the UV cutoff, we obtain the bound
\be
\Lambda_{\rm caus}\gtrsim \Lambda_{\rm UV}  \ \ \longrightarrow\ \  
r_5\gtrsim \sqrt{\frac{\Lambda_{\rm UV}}{\Lambda_{\rm IR}}}\,,
\label{thelastlimit}
\ee
that using  \eq{couplingsKK} can be rewritten as
\be
\kappa_5\, g_{\rm KK, grav}\,  g_{\rm KK, gauge}\lesssim
\frac{\Lambda_{\rm IR}}{\Lambda_{\rm UV}}\ll 1\,.
\label{limit5}
\ee
This result  looks quite surprising from the point of view of the 5D theory.
It tells us that either we have problems with causality  
($\Lambda_{\rm caus}\lesssim \Lambda_{\rm UV}$)
or  the  couplings of the KK spin-1 and spin-2 states  are not free parameters but related by \eq{limit5}.
Interestingly,  \eq{thelastlimit} can be satisfied
  when the  5D parameters are matched with the   ones expected from the dual description:\footnote{This is equivalent
to recall that    KK gravitons are dual to  glueballs,  while KK gauge bosons are dual to mesons, and then
 we have
$\kappa_5\, g_{\rm KK, grav}\,  g_{\rm KK, gauge}
\sim \kappa_g\, g_{\rm glueball}\,  g_{\rm meson}
\sim {1}/{\sqrt{N_c}}\ll 1$ where  in the last step  we used \eq{largeNcouplings}.}
\be
(M_{P5}L)^3\sim N_c^2 \ ,\ \ \    M_{5}L\sim N_c\ ,\ \ \ \kappa_{5}\sim N_c
 \label{scaling54}
\,,
\ee
that leads to  $r_5\sim  N_c^{1/4}\to\infty$  in the large-$N_c$ limit.

In the case in which the model  also contains a massless  dilaton (radion), that as we said requires    some fine-tuning \cite{Randall:1999ee}, one  finds  a bound on  its coupling to two gravitons similar to \eq{newbound}:
$|\kappa_{\rm dilaton}|\lesssim {M_P F_{\rm dilaton}}/{\Lambda^2_{\rm caus}}$. Using
\eq{holorelations} and $F^2_{\rm dilaton}\sim (M_{P5}L)^3\Lambda_{\rm IR}^2$, one gets
\be
\frac{|\kappa_{\rm dilaton}|}{(M_{P5}L)^3}\lesssim \frac{\Lambda_{\rm UV} \Lambda_{\rm IR}}{\Lambda_{\rm caus}^2}\,.
\label{dilat}
\ee
If higher-spin states  are infinitely heavy,  \eq{dilat} tells us that  $\kappa_{\rm dilaton}=0$.
This is  in fact  what one  finds  from the explicit calculation  using \eq{lagrangian5}.
 A coupling of the dilaton to two gravitons can only be generated if higher-dimensional terms are included
in \eq{lagrangian5}.

\subsubsection{5D  models in a compact flat space}

Similarly,  a model of  weakly-coupled resonances can be obtained  from the 5D model of \eq{lagrangian5}
in flat space with  the extra dimension compactified  in a segment of length $L$.
We get in this case  \cite{Barbieri:2003pr}
\be
F^2_\pi\sim M_5/L\ , \ \  M_P^2 \sim M^3_{P5}L\,,
\ee
and  from \eq{cutoff}
\be
\Lambda^4_{\rm caus}\sim \frac{M_5 M^3_{P5}}{\kappa_5^2}\,.
\label{flat}
\ee
Note that since $\Lambda_{\rm caus}$ in \eq{flat}  does not depend on $L$, 
this should  be  a causality cutoff for any 5D model defined by \eq{lagrangian5}, independently
of the spacetime metric.
In fact,  we could have derived  this bound  directly in 5D  rather than in 4D.

We can again ask what conditions the 5D parameters must satisfy in order to have
$\Lambda_{\rm caus}$ above the  cutoff scale of the 5D theory that, based on naive perturbativity,  is given by 
$\Lambda_5\sim {\rm Min}[M_5,M_{P5}]$. 
We  see that  (taking for concreteness $M_{P5}<M_5$)  
\be
\Lambda_{\rm caus}\gtrsim \Lambda_5\ \ \longrightarrow\ \ |\kappa_5|\lesssim \left(\frac{M_5}{ M_{P5}}\right)^{3/2}\,.
\label{newresult}
\ee
Surprisingly, this bound is not consistent with the scaling \eq{scaling54} in the large-$N_c$ limit.
This implies that either   new physics (states with $J\geq 4$)   appears below the cutoff $\Lambda_5$,
or the 5D parameters must depart from the simple scaling \eq{scaling54} to satisfy \eq{newresult}. 

It is interesting to check what happens in explicit holographic models derived from string theory.
Let us consider the  model of \cite{Kruczenski:2003be}, based on a $D3/D7$ system.    One has
\be
M^3_{P5}\propto N_c^2\ , \ \ 
M_{5}\propto N_c\ , \ \ 
\kappa_{5}\propto N_c\ , \ \ 
M_{\rm string}\propto \lambda^{1/4}\,,
\label{stringy}
\ee
where $M_{\rm string}$ is the string scale.
Using \eq{stringy} in \eq{flat}, one finds 
\be
\Lambda_{\rm caus}\propto N_c^{1/4}\,.
\ee
Consequently,  for small string coupling $g_{\rm string}=g^2_{\rm YM}/4\pi\lesssim 1$, where
$\lambda\lesssim N_c$, this model predicts   that $M_{\rm string}\lesssim \Lambda_{\rm caus}$, i.e.,
states with $J\geq 4$  indeed appear   below the causality bound.

\section{Conclusions}
\label{sec:conclude}

We have studied the implications of  causality and unitarity    on theories with 
a pseudo-Goldstone $\eta$ coupled to  two  gravitons  like in   \eq{etahh}, 
an interaction that arises from  $U(1)$-gravitational anomalies.
We  first considered  gravity as  an external  source probing the theories,
and  showed  that  no constraint can be obtained in this case on the anomaly coefficient
$\kappa_g$.

Therefore, we  needed to couple the  theory  to dynamical gravitons,
and consider dispersion relations  with  $1/t$ poles  arising from the graviton exchange.
Integrating these dispersion relations    over $t$ 
weighted by some  particular functions to  guarantee  positivity, we have   been able to 
derived bounds on $\kappa_g$.
   In particular,
 from the graviton-graviton scattering amplitude, we have derived
\eq{firstbound},  that can  be interpreted  as an upper bound on the mass of  $J\geq 4$ states.
Similarly, 
from the eta-graviton scattering amplitude, we obtained the  bound
\eq{thirdbound} that provided an upper bound on the  mass of $J\geq 2$ states.

These upper bounds  on the mass scale at which new states must  appear  
can be considered as a cutoff scale for the EFT of  $\eta$ and the  graviton.
 For axion EFTs we have seen that we expect this  cutoff to be 
at  $\Lambda_{\rm caus}\sim\sqrt{M_P f_a}$.

For  strongly-coupled gauge theories in the large-$N_c$ limit, that can be described as 
 models of massive higher-spin resonances, 
the cutoff scale $\Lambda_{\rm caus}$  was found to be larger than  the quantum gravity scale  due to the presence of glueballs whose couplings scale differently from that of mesons, as shown in \eq{largeNcouplings}.
This  happens if  the number of fermions is much smaller than $N_c$ and the  ’tHooft coupling $g_{\rm YM}^2 N_c$ is not large. 
For large ’tHooft coupling, however, where these theories can have an holographic 5D dual, we have seen that 
$\Lambda_{\rm caus}$ can lie below any other cutoff scale  (based on perturbativity),
leading to  consistency conditions on the parameters of the 5D models, 
as  for example \eq{limit5} and \eq{newresult}.

Our analysis for  4D theories with $U(1)$-gravitational anomalies deserves  future research.
For example, one could try to
improve the bounds by a better choice of the smearing functions, 
explore   UV completions beyond tree-level, or extend the  analysis to theories in higher dimensions.
We leave all this for  future work.

\vspace{1.0cm}

\section*{Acknowledgments}
We are very grateful to Francesco Riva and Brando Bellazzini for  valuable discussions.
We also thank Francesco Riva for a critical reading of the manuscript.
This work   has been  partly supported by the research grants 2021-SGR-00649 and 
 PID2023-146686NB-C31 funded by MICIU/AEI/10.13039/501100011033/ and by FEDER, UE.
T.M is partly supported by the Yan-Gui Talent Introduction Program (grant No. 118900M128) and Chinese Academy of Sciences Pioneer Initiative "Talent Introduction Plan".

\vspace{1.0cm}

\appendix

\section{Positivity in smeared dispersion relation at large $J$}
\label{applargeJ}
Here we show  which conditions the smearing functions must satisfy 
in order to  guarantee  positivity conditions as  \eq{positivity}
in the limit  $J\to \infty$. 
In particular, we will study the smeared  Wigner $d$-function appearing in \eq{aubu} that we generically write as (changing variables $t\to-t$)
\be
F(m,J)=\int_0^{|t|_\text{max}} f(t)\, d_{h,h}^J(1-2t/m^2) dt \, .
\label{smdr}
\ee
Here $f(t)$ refers   either to the smearing function,  $A(-t)$ or $B(-t)$,   or the quotient 
 $A(t)/(m^2-t)^3$ or $B(t)/(m^2-t)^4$ appearing in \eq{aubu}.

In the limit $J\to \infty$ the Wigner $d$-functions have a well-known behaviour given by the Darboux formula
\be
d_{h,h}^J(1-2t/m^2) \underset{J\rightarrow\infty}{\longrightarrow} \sqrt{\frac{m^2}{\pi J \sqrt{(m^2-t)t}}}\cos\Big(2(J+1/2) \sin^{-1}(\sqrt{t/m^2}) -\pi/4\Big)\, ,
\label{dblim}
\ee
which in turn we can expand in the regimes $t/m^2\to 0$ and $t/m^2\to1$, giving
\bea
d_{h,h}^J(1-2t/m^2) \underset{J\rightarrow\infty}{\longrightarrow} 
\begin{cases} 
J_{0}\big(2\sqrt{\mathcal{J}^2t/m^2}\big) + O({J}^{-3/2}) \hspace{3.9cm}
 t/m^2 \to 0\, , \\[12pt]
(-1)^{h+J} J_{2h}\big(2\sqrt{\mathcal{J}^2(1-t/m^2)}\big) + O({J}^{-3/2})  \hspace{1.2cm}
 t/m^2 \to 1\, ,
\end{cases}
\label{asD}
\eea
where we defined $\mathcal{J}^2 = J(J+1)$. 
Note that the limits above are correct as long as $t/m^2$ goes to zero (or $1$) slower than $1/J$, otherwise the expansion in $t/m^2$ and the one in $J$ do not commute (first term in $J$ becomes of the same order as the second and so on). With these limits in mind we can go back to the smeared dispersion relation \eq{smdr}. 

Let us initially consider the case where $m^2=|t|_\text{max}=M^2$. We can divide the integration region in three parts
and  change  coordinates $t=p^2$:
\be
F(M,J)=
\bigg(\int_0^{a} +  \int_a^{M-b} + \int_{M-b}^M \bigg) f(p) d_{h,h}^J(1-2p^2/M^2) p dp \, ,
\label{ints}
\ee
We choose $a,b$ to be zeros of the Wigner $d$-functions. 
This assures that if  $f(p)$ is a regular function, the middle integral in \eq{ints} goes to zero in the large $J$ limit.
This happens because we convolute the  rapidly oscillating Wigner $d$-function  with a regular function, such that the integral averages out to zero. 
We are left then with the integration regions between $(0,a)$ and $(M-b,M)$. 
By taking $a$ and $b$  small,\footnote{Due to the rapidly oscillating behaviour of the Wigner $d$-function at large $J$, we can always find a zero  at $p$ close to 0 and $M$.} we can use \eq{asD} and 
analytically perform the first and third  integration in \eq{ints}  after Taylor expanding $f(p)$ for $p$ close to zero and $M$ respectively. 
We find 
\be
F(M,J) \underset{J\rightarrow\infty}{\longrightarrow}  \bigg(f'(0)\frac{1}{8J^3}+O(J^{-4})\bigg) + (-1)^{h+J}\bigg(f(M)\frac{h}{2J^2}+O(J^{-4})\bigg)\, .
\ee
By demanding $f(M)=0$, we assure that the oscillating term in $J$ becomes $O(1/J^4)$ and 
is subdominant compared to the non-oscillating term, 
as long as $f'(0)\not= 0$.
We then get
  $F(M,J)\to\  f'(0)/J^3$  at large $J$.
This behaviour is also observed in our numerical results shown in Fig.~\ref{fig:asbsaubu}. 

We conclude then that at large $J$ and for $m=M$, positivity is guaranteed if 
$f(M)=0$ and $f'(0)>0$. These conditions  determined our smearing functions in \eq{functions} and \eq{cc}.

For the case $m^2>|t|_\text{max}$,
one can repeat the same procedure as above, with the exception that we cannot use 
the expansion \eq{asD} for $t/m^2\to 1$. One gets again that $F(m,J)$ drops at large $J$ as $\sim1/J^3$ with definite sign.

\section{Alternative bound on $\kappa_g$ from graviton scattering}

\label{2ndbound}

An alternative bound to \eq{firstbound} can be derived in the following way.
Let us consider the smeared dispersion relation \eq{drmp2} and \eq{drkappa2} and define
 \be
\hat A_{s,u}(m_i,J_i)\equiv A_{s,u}(m_i,J_i)\, m_i^4\ \ \  {\rm and}\ \ \ 
\hat B_{s,u}(m_i,J_i)\equiv B_{s,u}(m_i,J_i)\, m_i^6\,.
 \ee
Using  the trigonometric (Cauchy-Schwartz) inequality 
\be
 \big(\sum_i a_i b_i\big)^2 \leq \big(\sum_i a_i a_i\big) \big(\sum_i b_i b_i\big)\, ,
\label{trineq}
\ee
we   can write 
\bea
\bigg(\sum_i \frac{\hat B_s(m_i,J_i)}{m_i^6} g_{++,i}^2 
+ \sum_j \frac{\hat B_u(m_j,J_j)}{m_j^6} g^2_{+-,j} \bigg)^2\leq\hspace{4cm} \nonumber \\
 \bigg(\sum_i \frac{(\hat B_s(m_i,J_i))^2}{m_i^4}g_{++,i}^2+\sum_j \frac{(\hat B_u(m_j,J_j))^2}{m_i^4}g_{+-,j}^2\bigg) \bigg(\sum_i \frac{g_{++,i}^2}{m_i^8}+\sum_j \frac{g_{+-,j}^2}{m_i^8}\bigg)\,.
 \label{ineqtrig}
\eea
The two terms on the RHS can be related to other physical quantities as follows.
First, demanding to the smearing functions to satisfy
\be
(\hat B_s(m,J))^2 \leq \hat A_s(m,J) \ \ \text{and} \ \ (\hat B_u(m,J))^2 \leq  \hat A_u(m,J)\,,
\label{conditionlh}
\ee
we have 
\be
\bigg(\sum_i \frac{\hat B_s(m_i,J_i)^2}{m_i^4}g_{++,i}^2+\sum_j \frac{\hat B_u(m_j,J_j)^2}{m_i^4}g_{+-,j}^2\bigg)\leq 
\bigg(\sum_i \frac{\hat A_s(m_i,J_i)}{m_i^4}g_{++,i}^2+\sum_j \frac{\hat A_u(m_j,J_j)}{m_i^4}g_{+-,j}^2\bigg)\,,
\label{onemore}
\ee
where the RHS can be identified with the RHS of \eq{drmp2}.
Secondly,  using \eq{drcase1} for  $t\rightarrow0$, 
we obtain
\be
b_{0,0}=\sum_i \frac{g_{++,i}^2}{m_i^8}+\sum_j \frac{g_{+-,j}^2}{m_i^8}\,.
\label{bdef}
\ee
Now, using \eq{onemore} and \eq{bdef} together with \eq{drmp2} and  \eq{drkappa2},
we can rewrite  \eq{ineqtrig} as 
 \be
\bigg(\int_0^{|t|_\text{max}}dt \bigg(\frac{\kappa_g^2}{M_P^4 F_\pi^2} +\frac{\kappa_3^2}{ M_{P}^6}t^2\bigg) B(-t)\bigg)^2
\leq
\frac{b_{0,0}}{M_P^2}\int_0^{|t|_\text{max}} dt \frac{A(-t)}{t} \,.
\label{boundUV}
\ee

 Two functions $A(t)$ and $B(t)$ that guarantee \eq{conditionlh} to be satisfied are
\be
A(t) =B(t) =\frac{1}{|t|_\text{max}}\left(1-\sqrt{-{t}/{|t|_\text{max}}}\right)^4\,.
\label{abf}
\ee
Plugging the above functions  into \eq{boundUV}
and   taking $|t|_\text{max}=M^2_u$,
 we obtain the bound \eq{boundwilson}.

\section{Bounds from $ \eta h^+  \rightarrow h^+ h^+$  and the eikonal limit}
\label{elimit}

We can also use the process $ \eta h^+  \to h^+ h^+$ to obtain a bound on $\kappa_g$.
At  low-energy the amplitude for this process is given by
\be
{\cal M}_{\eta+--} \to s^3 t u\bigg( \frac{\kappa_g}{M_{P}^3F_\pi}\frac{1}{stu}+\frac{\kappa_3\kappa_g}{M_{P}^5F_\pi s}+e_{0,0}+\dots\bigg)\,,
\label{eq:epmm}
\ee
where the first term comes from the graviton exchange and $e_{0,0},...$  are  Wilson coefficients.
From the $k=2$ dispersion relation in the complex $s$-plane at fixed $t$, one obtains a sum rule for $\kappa_g$:
\be
\frac{\kappa_g}{M_P^3F_\pi}
-\frac{\kappa_3\kappa_g}{M_{P}^5F_\pi}t^2=\bigg\langle \frac{d_{-2,0}^J(1+2t/m^2)}{m^4}  \bigg\rangle^{g_{\eta+}g_{--}}+\bigg\langle \frac{m^2d_{2,4}^J(1+2t/m^2)}{(m^2+t)^3}  \bigg\rangle^{g_{\eta-}g_{-+}}\,, 
\label{bsmearing}
\ee
where, as in  \eq{drmp},  we must be aware that  the $t$-expansion is not convergent. Indeed, expanding in $t$ one finds that  the LHS is $O(t^0)$ while the RHS is $O(t)$. We therefore need to smear this sum rule. Using  $D(t)$ as a smearing function we have 
\be
\int_{-|t|_\text{max}}^{0} dt\ D(t) \bigg(\frac{\kappa_g}{M_P^3F_\pi}-\frac{\kappa_3\kappa_g}{M_{P}^5F_\pi}t^2\bigg)=\sum_i D_s(m_i,J_i) g_{\eta+,i}g_{--,i} + \sum_j D_u(m_j,J_j) g_{\eta-,j}g_{-+,j}\, ,
\label{irleft}
\ee 
where
\bea
D_s(m,J)&=& \int_{-|t|_\text{max}}^0 D(t)\ \frac{d_{-2,0}^J(1+2t/m^2)}{(m^2)^2}\ dt \nonumber \, ,\\
D_u(m,J)&=& \int_{-|t|_\text{max}}^0 D(t) \frac{m^2d_{2,4}^J(1+2t/m^2)}{(m^2+t)^3}\ dt \, .
\label{xiu}
\eea
\eq{irleft} relates the  anomaly coefficient $\kappa_g$  to  sums involving products of different meson couplings that can have either  sign. 
Therefore, as in Appendix \ref{2ndbound}, we will need to use the trigonometric  inequality
\eq{trineq}
to relate the RHS of \eq{irleft} to the elastic process  $\cal M_{--++}$ (A2 in Fig.~\ref{fig:diagrams1}) 
and $\cal M_{\eta+\eta-}$ (B in Fig.~\ref{fig:diagrams2})
whose dispersion relations were already detailed in Sec.~\ref{gravgrav} and Sec.~\ref{etagrav} respectively.

 Using the trigonometric inequality \eq{trineq}, we have 
 \bea
&&\bigg(\sum_i \sqrt{C_s(m_i,J_i)}\sqrt{A_s(m_i,J_i)} g_{\eta+,i}g_{--,i} + \sum_j \sqrt{C_s(m_i,J_i)}\sqrt{A_s(m_i,J_i)} g_{\eta-,i}g_{-+,i} \bigg)^2\leq\nonumber \hspace{1.15cm}\\
&& \bigg(\sum_i C_s(m_i,J_i) g_{\eta+,i}^2+\sum_j C_u(m_j,J_j) g_{\eta-,j}^2\bigg)  \bigg(\sum_i A_s(m_i,J_i) g_{--,i}^2+\sum_j A_u(m_j,J_j) g_{-+,j}^2\bigg)\nonumber\\
&&= 
 \bigg(\int_0^{|t|_\text{max}}C(-t) \bigg(\frac{1}{M_P^2t} +\frac{\kappa_g^2 t}{M_P^4F_\pi^2} \bigg)dt\bigg) \bigg( \int_0^{|t|_\text{max}}A(-t)\frac{1}{M_P^2t} dt\bigg)
 \,,
 \label{ineqtrig2}
\eea
where for the last expression we have used \eq{drmp2} and \eq{smdispelastic} and  parity to relate the couplings $g^2_{--,i}=g^2_{++,i}$.

Demanding the smearing functions to satisfy the following set of relations\footnote{This obviously demands $A_{s,u}$ and $C_{s,u}$ to be positive. 
Notice that  $D_u$ is  non-zero only for $J_i\geq 4$.}
\bea
D_s(m_i,J_i) \leq \sqrt{C_s(m_i,J_i)}\sqrt{A_s(m_i,J_i)}\hspace{1cm} \forall m^2_i\geq M_s^2 \ , \ \forall J_i\geq 2 \, , \label{constraintssxi} \\
D_u(m_i,J_i) \leq \sqrt{C_u(m_i,J_i) }\sqrt{A_u(m_i,J_i)}\hspace{1cm} \forall m^2_i\geq M_u^2 \ , \ \forall J_i\geq 2 \, ,
\label{constraintsxi}
\eea
it can be  guaranteed that  the squared of the RHS of  \eq{irleft}  is smaller than the LHS of \eq{ineqtrig2}, giving the following condition:
\bea
&&\bigg[\int_{0}^{|t|_\text{max}} dt\ D(-t) \frac{\kappa_g}{M_P^3F_\pi}\left(1-\frac{\kappa_3t^2}{M_p^2}\right)\bigg]^2\leq
\nonumber\\
&&\bigg(\int_0^{|t|_\text{max}}C(-t) \bigg(\frac{1}{M_P^2t} +\frac{\kappa_g^2 t}{M_P^4F_\pi^2} \bigg)dt\bigg) \bigg( \int_0^{|t|_\text{max}}A(-t)\frac{1}{M_P^2t} dt\bigg)\, .
\label{boundirxi}
\eea
Notice that $\kappa_g^2$ appears in both sides of \eq{boundirxi}. 
Therefore a non-trivial bound on $\kappa_g^2$ can only be obtained if the smearing functions also satisfy
\be
\bigg(\int_{0}^{|t|_\text{max}}dt\ D(-t)\bigg)^2\geq\bigg(\int_0^{|t|_\text{max}}C(-t) t  dt\bigg) \bigg( \int_{0}^{|t|_\text{max}}A(-t)\frac{1}{t} dt\bigg)\,,
\label{conditionextra}
\ee
that makes the $\kappa_g^2$ contribution of the RHS of \eq{boundirxi} smaller than that of the LHS.

A choice of functions which satisfies all the above constraints is
\bea
D(t)&=& J_2(\sqrt{-\beta^2 t/|t|_\text{max}}) \left(1-\sqrt{-t/|t|_\text{max}}\right)^4 \, , \nonumber \\
C(t)=A(t)&=& J_0(\sqrt{- \beta^2t/|t|_\text{max}}) \left(1-\sqrt{-t/|t|_\text{max}}\right)^4 \, ,
\label{smfuncs}
\eea
where $\beta$ is a dimensionless parameter and $J_n$ are Bessel functions.
 These latter are chosen to make  apparent   the eikonal limit, as we will discuss below.
As in Sec.~\ref{etagrav}  we take  $|t|_\text{max}=M^2_2$ where $M_2$ is the lightest $J\geq 2$ state with $g_{\eta\pm,i}\not=0$.

It is simple to verify that the inequalities \eq{constraintssxi} and \eq{constraintsxi} are satisfied for the functions of
\eq{smfuncs} for  any choice of $\beta^2\geq0$. 
One can also check that  \eq{conditionextra} is satisfied for  $\beta\geq6$ that 
makes $\int_0^{M_2^2}C(-t) t  dt\leq 0$.
We find that the value of $\beta$ which maximizes the bound is $\beta\sim 8$, giving \eq{secondbound}.

\subsection{Recovering the Eikonal Limit}
\label{eiksub}
In the limit   $\beta \to \infty$, we can identify $\beta$ with the impact parameter of the scattering in the eikonal limit.
To see this, let us consider, as an example, $D_u$ defined in \eq{xiu}  with the smearing function
\be
D(t)=
e^{-\frac{\alpha}{\beta^2}  t} J_2(\beta\sqrt{- t/M_{2}^2}) \left(1-\sqrt{-t/M_{2}^2}\right)^4 \,,
\ee
where, with respect to \eq{smfuncs},  we have introduced an exponential factor  ($\alpha$ is a parameter to be fixed later) to regularize  the   integral over $t$  in the $\beta\to \infty$ limit.\footnote{This is to suppress  the oscillatory behaviour of  $J_2(\beta\sqrt{- t/M_2^2})$  when taking $\beta\to\infty$.}
We have also taken $|t|_\text{max}=M_{2}^2$.
After a change of variables $t'=-\beta^2 t$,  we have
\be
D_u(m,J)= \frac{1}{\beta^2}\int^{\beta^2 M_2^2}_0 e^{-\alpha t'}\left(1-\sqrt{t'/(\beta^2 M_2^2)}\right)^4 
{J}_2(\sqrt{t'/ M_2^2})\, \frac{d_{2,4}^J(1-2t'/(\beta m)^2)}{(1-t'/(\beta m)^2)^3}\ dt'\, .
\label{limitep}
\ee
For $\beta\to\infty$ the Wigner $d$-functions 
tends to  Bessel functions:
\be
d_{\lambda_1,\lambda_2}^J(1-2t'/(\beta m)^2)\  \ {\underset{\beta\rightarrow \infty}{\longrightarrow}}\ \ J_{|\lambda_1-\lambda_2|}\bigg(\frac{2\mathcal{J}\sqrt{t'}}{\beta m}\bigg)\, \ \ \ \text{with\quad } \mathcal{J}=\sqrt{J(J+1)}\,,
\ee
that using it  in \eq{limitep}  gives
\be
D_u(m,J)\ \ \underset{\beta\rightarrow\infty}{\longrightarrow} \ \ \frac{1}{\beta^2}\int_0^\infty e^{-\alpha t'}   {J}_2(\sqrt{ t'/M_2^2})  {J}_2\bigg(\frac{2\mathcal{J} \sqrt{t'}}{\beta m}\bigg)dt'\, .
\ee
If we now take the limit $\alpha \rightarrow 0$ the exponential term drops off and the expression becomes an integral of two Bessel  functions which satisfy an orthogonality condition (reason for which we chose $J_2$ for $D(t)$ and $J_0$ for $C(t)$ and $A(t)$ in \eq{smfuncs}) that gives
\be
D_u(m,J) \ \  \overset{\beta\rightarrow\infty}{\underset{\alpha\rightarrow0}{\longrightarrow}} \ \ \frac{M_2}{\beta^2}\, \delta\bigg(\frac{1}{M_2}-\frac{2\mathcal{J}}{\beta m}\bigg)\, .
\label{deltaresult}
\ee
This delta function will enter in the infinite sum over masses and spins in \eq{irleft}, selecting those  with  $2\mathcal{J}/m = \beta/ M_2$. This procedure can be repeated for $D_s, C_{s,u}$ and $A_{s,u}$, all of which will give similar result as the one in \eq{deltaresult}, which guarantees that the inequalities in \eq{constraintsxi} are satisfied. 
The  integrals 
in \eq{boundirxi} containing    $\kappa_g$  become 
\bea
\int^{M_2^2}_{0} dt\ D(-t)\frac{\kappa_g}{M_P^3F_\pi} &{\underset{\beta\rightarrow \infty }{\longrightarrow}}& \frac{\kappa_g}{\beta^2 M_P^3F_\pi} \int_0^{\infty} dt' J_2(\sqrt{t'/M_2^2}) e^{-\alpha t'}
\ \ \ {\underset{\alpha\rightarrow 0 }{\longrightarrow}} \ \ \ 
\frac{2M^2_{2}\kappa_g}{\beta^2M_P^3F_\pi} 
 \nonumber \, ,\\
 \int^{M_2^2}_{0} dt\ D(-t)\frac{\kappa_g\kappa_3t^2}{M_P^5F_\pi} &{\underset{\beta\rightarrow \infty }{\longrightarrow}}& \frac{\kappa_g\kappa_3}{\beta^6 M_P^5F_\pi} \int_0^{\infty} dt' J_2(\sqrt{t'/M_2^2}) e^{-\alpha t'}t'^2
\ \ \ {\underset{\alpha\rightarrow 0 }{\longrightarrow}} \ \ \ 
0
 \nonumber \, ,\\
\int_{0}^{M_2^2} dt\ C(-t)\frac{\kappa_g^2 t}{M_P^4F_\pi^2}
&{\underset{\beta\rightarrow \infty }{\longrightarrow}}&
\frac{\kappa_g^2}{\beta^4 M_P^4F_\pi^2}
 \int_0^{\infty} dt' J_0\big(\sqrt{t'/M^2_{2}}\big)e^{-\alpha t'} t' 
\ \ \ {\underset{\alpha\rightarrow 0 }{\longrightarrow}} \ \ \ 
0 \,.
\eea
The integrals containing   $1/M_P^2$ in \eq{boundirxi} have also  in the limit $\beta\to\infty$ and $\alpha\to 0$ 
a  logarithmic divergence  giving  
$\int_0^{\infty} dt' J_0\big(\sqrt{t'/M_2^2}\big) /t'\sim  \ln(M_2^2/M^2_\text{IR})$ 
(we have redefined $M_\text{IR}$ to absorb  a $\beta^2$ term). 
Using the above results   into \eq{boundirxi}, one finds the non-trivial bound 
\be
\frac{M_2^2 \kappa_g}{\beta^2M_PF_\pi}\leq \ln\bigg(\frac{M_2}{M_\text{IR}}\bigg)\, ,
\label{boundeikonal}
\ee
valid for $\beta\gg 1$.  
This bound could also be found alternatively from  time-delays in  $\eta$-graviton scatterings with  $\beta/M_2$ being identified with the impact parameter \cite{Camanho:2014apa}.

\bibliography{AMP}
\bibliographystyle{apsrev}

\end{document}